\documentclass[twocolumn]{aa}

\usepackage{txfonts}
\usepackage{graphicx}
\usepackage{natbib}
\usepackage{mathrsfs}
\usepackage{url}
\bibpunct{(}{)}{;}{a}{}{,}

\begin{document}

\title{High-pressure, low-abundance water in bipolar outflows} 
\subtitle{Results from a Herschel-WISH survey
\thanks{{\it Herschel} is an ESA space observatory with science instruments
provided by European-led Principal Investigator consortia and with important
participation from NASA.}
}

\titlerunning{High-pressure, low-abundance water in bipolar outflows}

\author{M. Tafalla
\inst{1}
\and
R. Liseau
\inst{2}
\and
B. Nisini
\inst{3}
\and
R. Bachiller
\inst{1}
\and
J. Santiago-Garc\'{\i}a
\inst{4}
\and
E. F. van Dishoeck
\inst{5,6}
\and
L. E. Kristensen
\inst{5}
\and
G. J. Herczeg
\inst{6,7}
\and
U. A. Y{\i}ld{\i}z
\inst{5}
}

\institute{Observatorio Astron\'omico Nacional (IGN),
Alfonso XII 3, E-28014 Madrid, Spain
\and 
Department of Earth and Space Sciences, Chalmers University of Technology, 
Onsala Space Observatory, 439 92 Onsala, Sweden
\and
INAF - Osservatorio Astronomico di Roma, via di Frascati 33, 00040 
Monte Porzio Catone, Italy
\and
Instituto de Radioastronom\'{\i}a Milim\'etrica (IRAM), Avenida Divina
Pastora 7, N\'ucleo Central, 18012 Granada, Spain
\and
Leiden Observatory, Leiden University, P.O. Box 9513, 2300 RA
Leiden, The Netherlands
\and
Max-Planck Institut f\"ur Extraterrestrische Physik (MPE),
Giessenbachstr. 1, D-85748 Garching, Germany
\and
Kavli Institute for Astronomy and Astrophysics, Peking University,
Yi He Yuan Lu 5, Hai Dian Qu, 100871 Beijing, P.R. China.
}

\offprints{M. Tafalla, \email{m.tafalla@oan.es}}
\date{Received -- / Accepted -- }

\abstract
{Water is a potential tracer of outflow activity due to its
heavy depletion in cold ambient gas and its copious 
production in shocks.
}
{We present a survey of the water emission in a sample of
more than 20 outflows from low mass young stellar objects with
the goal of characterizing the physical and chemical conditions
of the emitting gas. 
}
{We have used the HIFI and PACS instruments on board the Herschel Space
Observatory to observe the two fundamental lines of ortho-water
at 557 and 1670~GHz. These observations were part of
the ``Water In Star-forming regions with Herschel''
(WISH) key program, and have been complemented with CO and H$_2$ data.
}
{The emission from water has a
different spatial and velocity distribution         
from that of the $J$=1-0 and 2-1 transitions of CO.
On the other hand, it has a similar spatial distribution to
H$_2$, and its intensity follows
the H$_2$ intensity derived from IRAC images.
This suggests that water traces the outflow gas at hundreds
of kelvins responsible for the H$_2$ emission,
and not the component at tens of kelvins
typical of low-$J$ CO emission.
A warm origin of the water emission is confirmed by
a remarkable correlation between the intensities of
the 557 and 1670~GHz lines, which also
indicates the emitting gas has a narrow range of excitations.
A non-LTE radiative transfer analysis shows that
while there
is some ambiguity on the exact combination of
density and temperature values, 
the gas thermal pressure $nT$ is constrained
within less than a factor of 2. The
typical $nT$ over the sample is $4\times 10^{9}$~cm$^{-3}$K,
which represents an increase of $10^4$ with respect to 
the ambient value.
The data also constrain within a factor of 2 
the water column density, which varies over the
sample from
$2\times 10^{12}$ to $10^{14}$~cm$^{-2}$.
When these values are
combined with H$_2$ column densities, the
typical water abundance is only $3\times 10^{-7}$,
with an uncertainty of a factor of 3.}
{Our data challenge current C-shock 
models of water production due to a combination of wing-line profiles,
high gas compressions, and low abundances.
}


\maketitle


\section{Introduction}

Bipolar outflows are ideal laboratories to
study the physics and chemistry of interstellar medium shocks.
They result from the interaction between a (still mysterious)
supersonic
wind launched by a protostar 
and the cold, extended gas cloud from which the protostar was born
\citep{bac96,arc07}.
Their rich physical and chemical structure has attracted
intense attention from  both theorists and observers. 
Emission from H$_2$ vibration-rotation transitions, for example,
reveals  shock-heated gas at hundreds or few thousand
kelvins \citep{gau76}, while systematic
abundance enhancements of species like SiO and CH$_3$OH
evidences a rich chemistry driven by a combination of gas-phase reactions
and dust shock disruption  \citep{van98}.
Both physical and chemical activity in outflows seem
correlated with protostellar youth, likely due to the
combined effect of outflow weakening with time and gradual
clearing of the protostellar envelope \citep{bon96,taf11}.
As a result, the study of
the physical and chemical activity of
outflows is not only of interest for 
understanding ISM shocks, but constitutes a necessary step
to elucidate the still-mysterious physics of 
star formation.

The H$_2$O molecule constitutes an exceptional tool 
for
studying both the physics and chemistry of the shocked gas in outflows.
H$_2$O has been found to be heavily depleted 
in the unperturbed gas of cold, star-forming regions 
\citep{ber02,cas12}, and at the same time, 
is predicted to be copiously produced under the type of shock
conditions expected in outflows \citep{dra83,kau96,ber98,flo10}.
These extreme properties make H$_2$O a highly 
selective tracer of outflow activity, and indeed, 
H$_2$O maser emission has long been used as an outflow
signpost, especially in high-mass star-forming regions \citep{gen77}.
Unfortunately, maser emission, the only radiation from the 
H$_2$O main isotopologue observable from the
ground, is a notoriously difficult tool
for estimating emitting-gas parameters, since by its nature, it is 
highly biased to gas having specific, maser-producing
physical conditions. To extract the full potential of H$_2$O
as an outflow tracer, observations of its
thermal emission are needed, and this requires the use of 
a space-based telescope.

The Infrared Space Observatory (ISO) provided the first 
systematic view of the thermal H$_2$O emission from outflows.
The combined low angular and spectral resolution 
of the ISO data made it difficult to
compare the observed  H$_2$O  emission with that 
of other tracers observable from the
ground, like the low-$J$ transitions of CO. Still, these
pioneer ISO observations revealed strong H$_2$O emission 
toward a number of young low-mass outflows
from both the ground and excited energy levels,
indicating that at least part of the H$_2$O emission
originates in relatively warm gas
\citep{lis96,nis99,gia01,ben02}. Velocity-resolved 
H$_2$O observations were made possible first
by the Submillimeter Wave Astronomy Satellite (SWAS)
and later by Odin. These two satellites
observed the fundamental line of ortho-H$_2$O
at 557~GHz with velocity resolutions 
better than 1~km~s$^{-1}$, revealing line profiles
with high velocity wings of clear outflow
origin \citep{fra08,bje09}. However, neither SWAS nor
Odin, with their several arcmin telescope beams, 
could resolve spatially the outflow emission, and these
observations provided only a global view
of the thermal emission from H$_2$O in outflows.

The Herschel Space Observatory \citep{pil10} has finally
provided the combination of angular and spectral resolutions needed to
study in detail the emission of H$_2$O in nearby outflows.
Herschel instruments can observe a variety 
of ortho and para 
H$_2$O lines, opening up H$_2$O studies to the
same multi-line type of analysis commonly used with
other molecular tracers.
To maximize this potential,
the ``Water In Star-forming regions with Herschel'' 
(WISH)\footnote{\url{http://www.strw.leidenuniv.nl/WISH/}}
key program pooled more than 400 hr of 
telescope time with the goal of using H$_2$O and
related molecules
to study both the physical and chemical conditions
of the gas in nearby star-forming regions \citep{van11}.
A specific subprogram of WISH is dedicated to 
study the H$_2$O emission
from low-mass outflows, which are the ones most likely
to show emission free from multiplicity and
additional energetic phenomena.
Due to the limited observing time available,
the outflow subprogram was split into three
parts with specific goals: (i) mapping
three selected outflows to study the spatial distribution of
H$_2$O, (ii) multi-transition
observations toward two positions of each mapped outflow 
to constrain the H$_2$O excitation, and 
(iii) a survey of short integrations toward  
about 20 outflows to accumulate a 
statistically-significant sample of H$_2$O observations.
Results from the mapping part of the program 
have been presented by
\citet{nis10a} for the L1157 outflow, \citet{bje12} for
the VLA1623 outflow, and \citet{nis13} for the L1448 outflow.
Preliminary work on the multi-transition analysis has been 
presented by \citet{vas12} for the L1157 outflow and 
\citet{san12} for the L1448 outflow. In this paper, we
report on the results of the
statistical study of outflows.
Additional results concerning outflow emission
from different subprograms of WISH
have been presented by \citet{kri11},
\citet{kri12}, and \citet{her12} toward protostellar
positions, and by \citet{bje11} toward the HH54 outflow region.
Detailed observations of the L1157 outflow by the
Chemical HErschel Surveys of Star forming regions (CHESS)
program can be found in \citet{lef10}, \citet{cod10},
\citet{ben12}, and \citet{lef12}.

\section{Observations}

The survey presented here was designed as a first
look at the H$_2$O emission from a large number of bipolar
outflows using a moderate amount of telescope
time (approximately 7 hours).
This required a compromise between sample size, line selection,
and sensitivity,
and led to a strategy based on the observation of the two fundamental
transitions of ortho-H$_2$O toward two
positions in about 20 outflows, using a typical integration
time of 300 seconds per transition.

\subsection{Target selection}

The survey target sample
consists of 22 outflows, of which 17 are believed to be driven by class 0
sources, 3 are associated with class I sources, and 2 have driving sources
of undetermined class (see Table~\ref{centers} for
central positions and Table~\ref{summary} for the targeted outflow positions).
Having a large fraction of class 0 sources was preferred because
the outflows from these sources tend to be 
the most energetic and ``chemically active'' 
\citep{bon96,taf11}, and were therefore 
expected to provide the highest rate of water detection.
Intentionally, the list of exciting sources had a large overlap 
with the target list of the {\em low-mass YSO} subprogram of
WISH, which studies the water emission from the envelopes of
low-mass protostars \citep{van11,kri12}. For most overlap sources,
we selected one bright position in each outflow
lobe generally well offset from the protostar,
using  as a guide published maps of emission
from CO, SiO, or H$_2$. For sources with no overlap,
we commonly chose the YSO as 
one of the survey targets, although the decision was made on a 
case-by-case basis taking into account the outflow geometry and
our expectation for the brightest H$_2$O emission peak.

Given the diverse set of literature
maps used to select the targets, 
our sample is not
biased in a simple systematic way. It clearly represents
a group of outflow positions likely to have strong H$_2$O emission,
but our use of different tracers 
(CO, SiO, H$_2$) and literature maps of different quality and resolution 
made the sample significantly heterogeneous.
As we will see below, the diverse nature of the sample 
became a significant advantage at the time of the analysis, because it 
increased the dynamic range of the observed intensities
and probed (often inadvertently) 
a variety of emitting regions, and 
not just the brighter H$_2$O peaks.

After the survey was finished, we noticed that  one target 
position had been erroneously associated with 
a bipolar outflow.  This position corresponds to SERSMM4-B,
and had been included in the sample because of the
strong SiO and CH$_3$OH detections reported by \citet{gar02}. Later
CO(3--2) observations by \citet{dio10b}, however, found no association of this
position with the SERSMM4 outflow or with any other outflow from the Serpens
cluster. To avoid contaminating our sample with a non-outflow
position, the data from SERSMM4-B have been
excluded from the analysis.

\begin{table}[h!]
\caption[]{Target outflows and central positions for offset calculation.$^{(1)}$}
\label{centers}
\centering
\begin{tabular}{l c c r@{.}l c c}
\hline
\noalign{\smallskip}
\mbox{Source} & \mbox{$\alpha(J2000)$}  &
\mbox{$\delta(J2000)$} &  \multicolumn{2}{c}{$V_{\mathrm{LSR}}$} &
\mbox{$T_{\mathrm{bol}}^{(2)}$} & \mbox{Vel.}  \\
 & \mbox{$(\;^h\; ^m\; ^s\;)$} & \mbox{$(\;^\circ\; '\; ''\;)$} &
  \multicolumn{2}{c}{\mbox{(km~s$^{-1}$)}} & \mbox{(K)} & \mbox{Ref.} \\
\noalign{\smallskip}
\hline
\noalign{\smallskip}
\mbox{N1333I2} & $03\;28\;55.6$ & $+31\;14\;37$ & 7&5 & 53 & (3) \\
\mbox{N1333I3} & $03\;29\;03.8$ & $+31\;16\;04$ & 7&5 & 136 & (3) \\
\mbox{N1333I4A} & $03\;29\;10.5$ & $+31\;13\;31$ & 7&2 & 34 & (3) \\
\mbox{HH211} & $03\;43\;56.8$ & $+32\;00\;50$ & 9&1 & 30 & (3) \\
\mbox{IRAS04166} & $04\;19\;42.6$ & $+27\;13\;38$ & 6&7 & 56 & (4)  \\
\mbox{L1551} & $04\;31\;34.1$ & $+18\;08\;05$ & 6&8 & 106 & (3)  \\
\mbox{L1527} & $04\;39\;53.9$ & $+26\;03\;10$ & 5&9 & 42 & (3) \\
\mbox{HH1-2} & $05\;36\;22.8$ & $-06\;46\;07$ & 9&4 & - & (5) \\
\mbox{HH212} & $05\;43\;51.4$ & $-01\;02\;53$ & 1&6 & 41 & (6) \\
\mbox{HH25MMS} & $05\;46\;07.3$ & $-00\;13\;30$ & 10&3 & 47 & (7) \\
\mbox{HH111} & $05\;51\;46.3$ & $+02\;48\;30$ & 8&7 & 69 & (8) \\
\mbox{HH46} & $08\;25\;43.9$ & $-51\;00\;36$ & 5&3 & 112 & (9) \\
\mbox{BHR71} & $12\;01\;36.3$ & $-65\;08\;53$ & -4&5 & 48 & (9) \\
\mbox{HH54B} & $12\;55\;50.3$ & $-76\;56\;23$ & 2&4 & - & (10) \\
\mbox{IRAS16293} & $16\;32\;22.8$ & $-24\;28\;36$ & 4&0 & 45 & (3) \\
\mbox{L483} & $18\;17\;29.9$ & $-04\;39\;39$ & 5&4 & 49 & (3) \\
\mbox{S68N} & $18\;29\;48.0$ & $+01\;16\;46$ & 8&8 & 45 & (3) \\
\mbox{SERSMM1} & $18\;29\;49.8$ & $+01\;15\;21$ & 8&5 & 39 & (3) \\
\mbox{SERSMM4} & $18\;29\;56.6$ & $+01\;13\;15$ & 8&1 & 33 & (3) \\
\mbox{B335} & $19\;37\;00.9$ & $+07\;34\;10$ & 8&3 & 42 & (3) \\
\mbox{N7129FIR2} & $21\;43\;01.7$ & $+66\;03\;24$ & 9&5 & 52 & (11) \\
\mbox{CEPE} & $23\;03\;13.1$ & $+61\;42\;26$ & -13&0 & 56 & (12) \\
\hline
\end{tabular}
\begin{list}{}{}
\item[] (1) All central positions as in \cite{van11} except S68N, which has
an offset of $9''$. See table~\ref{summary} for the offsets of
the observed positions; (2) Bolometric temperature as defined by
\citet{mye93} and estimated using data from Spitzer
telescope observations \citep{vel07,eva09,gut09,reb10}, AKARI \citep{ish10,yam10},
IRAS \citep{bei88,hur96}, and JCMT \citep{dif08};
(3) \citet{mar97}; (4) \citet{taf04}; (5) \citet{mar88};
(6) \citet{wis01}; (7) \citet{cho99}; (8) \citet{sep11}; (9) \citet{bou95};
(10) \citet{bje11}; (11) \citet{fue05}; (12) \citet{lef96}
\end{list}
\end{table}

\subsection{HIFI observations of H$_2$O(1$_{10}$--1$_{01}$)}

\begin{figure}
\centering
\resizebox{\hsize}{!}{\includegraphics{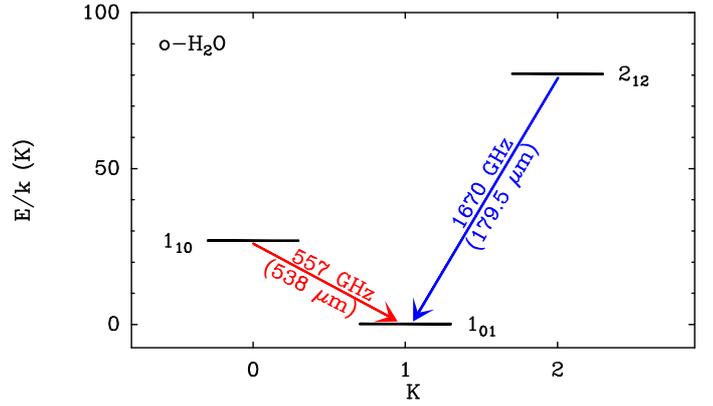}}
\caption{Lower part of the ortho-H$_2$O energy diagram
illustrating the 
two transitions observed in the outflow survey. The colors of the arrows
indicate the instruments used to observed the two transitions (red
for HIFI and blue for PACS), and the energies of the levels are referred to
the fundamental level of ortho-H$_2$O (instead of the frequently used
para-H$_2$O).
\label{levels}}
\end{figure}

We observed our target sources in
H$_2$O(1$_{10}$--1$_{01}$) 
(rest frequency 556.9360020 GHz, \citealt{pic98},
see Fig.~\ref{levels}) with HIFI \citep{deg10} 
between April 2010 and April 2011. These observations,
from now on referred to as the ``557~GHz'' observations,
were done initially in position switching (PS) mode
using a reference $10'$ or more away from the source.
Experience within the WISH project, however, showed
that dual beam switching (DBS) with a $3'$ chop 
produced flatter baselines than PS, and the 
observing mode was changed to DBS after the first set of data were obtained.
In all observations, the
local oscillator (LO) was tuned so that both H$_2$O(1$_{10}$--1$_{01}$) and
NH$_3$($JK=10$--00) (rest frequency 572.4981599 GHz, \citealt{pic98}) 
fell inside the bandpass. 
A few initial spectra had the NH$_3$ line (coming from the upper sideband) 
falling too close to the
H$_2$O line (coming from the lower sideband), and the LO was re-tuned in
the remaining observations to separate the lines and
avoid possible overlaps.

During the observations, both
the horizontal and vertical components of
the polarization were detected, 
and the Wide Band Spectrometer (WBS) and High Resolution Spectrometer (HRS)
were used to provide velocity resolutions of 0.6 and 0.13~km~s$^{-1}$,
respectively.
The data were calibrated using the Standard Product Generation (SPG) 
pipeline in HIPE v6.1 \citep{ott10}, and then converted to 
the GILDAS program
CLASS\footnote{\url{http://www.iram.fr/IRAMFR/GILDAS}} for
first order baseline subtraction, average of polarizations,
and further processing. 
According to in-flight calibration measurements, the telescope beam size
at 557~GHz was $39''$,
and the beam efficiency was 0.76.
\citep{roe12}.
All our intensities are expressed in $T_{\mathrm{mb}}$ units, with a
nominal uncertainty estimated as $<15$\%.

\subsection{PACS observations of H$_2$O(2$_{12}$--1$_{01}$)}

The observations of the H$_2$O(2$_{12}$--1$_{01}$) line
(rest frequency 1669.9047750 GHz, \citealt{pic98}, see Fig.~\ref{levels}) 
were carried out with the PACS instrument
\citep{pog10} between October 2009 and September 2011 
in line spectroscopy mode. This observing mode provided
a $5 \times 5$ array of velocity-unresolved spectra
(FWHM $\approx 200$~km~s$^{-1}$) covering a field of view 
of $47'' \times 47''$. Each spectrum represents a sample on 
a $9\farcs4 \times 9\farcs4$ pixel, which is slightly undersized
compared to the $13''$ telescope beam at the operating
frequency.
The observations, from now on referred to
as the ``1670~GHz'' observations, used the 
pointed chopping/nodding mode with a so-called large throw
of $6'$. 

Depending on the date of the observation, the data were
calibrated with HIPE versions 4, 5, or 6 
using the standard reduction pipeline and a calibration scheme consistent
among the HIPE versions. After that,
the data were converted 
into CLASS format for first-order baseline subtraction and further
analysis. 
To compare with the HIFI 557~GHz data, the
PACS intensity scale of the 1670~GHz observations
(Jy px$^{-1}$) was converted into an equivalent brightness temperature scale
using the relation $T_\mathrm{B}  (K) = 5.6\: 10^{-3}\; S_\nu$(Jy px$^{-1}$), which
assumes square $9.4''$ pixels and an emitting
region larger than the 13$''$ beam.
A number of tests were carried out to ensure consistency between the
calibration of PACS and HIFI data, including a  comparison of
intensities from objects observed in the 1670~GHz line with both
instruments as part of different WISH subprograms.
These and other tests carried out by
the WISH team suggest that the
uncertainty level of the PACS calibration is on the
order of 20\%.

\subsection{Complementary IRAM 30m CO observations}

Complementary observations of the Herschel targets were carried out
with the IRAM 30m telescope between 2-4 May 2008. 
The observations consisted of CO(1--0) and CO(2--1) 
on-the-fly maps centered on the Herschel target position and
covering a region $80''\times 80''$. 
Each mapping observation lasted about 15 minutes and 
was made in position-switching mode using the
reference position initially chosen for the Herschel observations. 
Additional frequency switched spectra of most reference positions
were taken to correct for possible contamination by
residual emission. For each line, the two orthogonal polarizations 
were observed simultaneously
and later averaged, and 
both the 1MHz filter bank and the VESPA 
autocorrelator were used as backends to provide velocity resolutions
between 0.2 and 2.6~km~s$^{-1}$.
Data reduction was carried out with the CLASS software, and the intensity 
scale of the spectra was converted to $T_\mathrm{mb}$ using the 
facility-recommended efficiencies.

\subsection{IRAC archival data}

The IRAC instrument is a four-channel camera on the
Spitzer Space Telescope that operates
simultaneously at 3.6, 4.5, 5.8, and 8.0~$\mu$m 
with bandwidths between 0.8 and 3.0~$\mu$m
(channels IRAC1 to IRAC4). It
produces diffraction-limited images with a PSF between $1\farcs6$ and 
$1\farcs9$ depending on the wavelength (see \citealt{faz04} for a 
full description of the instrument).
Over the years, IRAC has been used to observe most of
our target objects as part of different projects, 
and all archival images are available
at the Spitzer Heritage
Archive (SHA)\footnote{\url{http://sha.ipac.caltech.edu/applications/Spitzer/SHA/}}.
From this archive, we downloaded the Level 2 images of each target
as reduced with the S18.18 pipeline, and we
have used them to complement our H$_2$O analysis.

Being relatively broadband 
($\sim 25$~\%), the different IRAC channels are sensitive
to both continuum and line emission. Of particular interest for our study 
are the lines from H$_2$, which include 
$v$ = 1-0 $O$(5)-$O$(7) and $v$ = 0-0 $S$(4)-$S$(13).
In regions of shocked gas, these
lines often dominate over the continuum
contribution, making the
IRAC images good tracers of the H$_2$ emission \citep{rea06,neu08}.
As shown by \citet{rea06} and \citet{neu08}, 
the $v$ = 1-0 lines lie inside the IRAC1 channel, and the $v$ = 0-0 lines
are distributed over the four channels following a pattern of 
decreasing $S$ number with increasing wavelength. As a result of
this order, the H$_2$ lines with lowest energy lie inside the IRAC4 passband, 
and this makes channel 4 of particular interest for our analysis. 
While this channel can suffer from potential contamination by PAH
emission \citep{rea06}, the IRS spectra from \citet{neu09} show
that even bright low-mass outflows like L1157, BHR71, or L1448
present
negligible PAH features in the IRAC4 (6.5-9.5~$\mu$m) 
band.
A comparison between IRAC1 and IRAC4 images for
the objects of our sample shows no appreciable 
differences in the morphology of the emission, again suggesting that
PAH contamination is negligible.

\section{Overview of the survey results}

\begin{table*}
\caption[]{Survey positions and fit results.}
\label{summary}
\centering
\begin{tabular}{l c | c c | c c c | c c}
\hline
\noalign{\smallskip}
& & \multicolumn{2}{c|}{\mbox{PACS$^{(1)}$}}
& \multicolumn{3}{c|}{\mbox{HIFI$^{(2)}$}} & 
\multicolumn{2}{c}{\mbox{PACS-HIFI$^{(3)}$}}\\
\mbox{Position} & \mbox{Offset}$^{(4)}$  &
\mbox{$I$[1670~GHz]$_{\mathrm{peak}}$} & \mbox{Diam.} &
\mbox{$I$[557~GHz]} & \mbox{$V_{\mathrm{LSR}}$} &
 \mbox{$\Delta V$} & \mbox{log[$N$(H$_2$O)]} & \mbox{log($nT$)} \\
& \mbox{$('',\;'')$} & \mbox{(K km s$^{-1}$)} 
& \mbox{$('')$} &
\mbox{(K km s$^{-1}$)} & \mbox{(km s$^{-1}$)} & \mbox{(km s$^{-1}$)}  
& \mbox{(cm$^{-2}$)} & \mbox{(cm$^{-3}$ K)}\\
\noalign{\smallskip}
\hline
\noalign{\smallskip}
\mbox{N1333I2-B} & $(-103, +23)$ & 3.5 (0.6)  &  30 (4.8) &  
6.34 (0.04) &  -0.14 (0.04) & 11.3 (0.08) & 13.4 (0.2) & 9.2 (0.1) \\
\mbox{N1333I2-R} & $(+67, -17)$  & 5.5 (0.8)  &  22 (2.3) & 
7.15 (0.06) &  14.0 (0.07) & 14.2 (0.1) & 13.3 (0.2) & 9.4 (0.1)  \\
\mbox{N1333I3-B2} & $(+20, -50)$  & 1.7 (0.2)  &  27 (3.1) &
2.65 (0.07) & 2.0 (0.3) & 20.5 (0.6) & 12.8 (0.2) & 9.4 (0.1) \\
\mbox{N1333I3-B1} & $(+20, -20)$ & 5.9 (1.4) &  18 (2.9) &
5.35 (0.08) & -4.6  (0.2)  &   25.4 (0.4) & 13.1 (0.1) & 9.7 (0.1) \\
\mbox{N1333I4A-B} & $(-6, -19)$ &  14.0 (2.9) &  34 (6.8) &
15.0 (0.1) &  1.3 (0.06) & 17.4  (0.1) & 13.5 (0.1) & 9.1 (0.1) \\
\mbox{N1333I4A-R} & $(+13, +25)$ & 9.3 (1.2) &  28 (3.5) &
 7.76  (0.07) & 13.3 (0.09) & 19.3 (0.2) & 13.4 (0.1) & 9.7 (0.1) \\
\mbox{HH211-R} & $(-37, +15)$ & 2.8 (0.6) &  28 (5.8) &
- & \mbox{No~data}  & - & - & - \\
\mbox{HH211-C} & $(0, 0)$ & 11.7 (1.6)  &  17  (1.7) &
6.53 (0.02) &  11.5  (0.07) &  41.4 (0.1) & 13.2 (0.1) & 9.8 (0.1)  \\
\mbox{HH211-B} & $(+37, -15)$ & 7.7 (1.0) &  14 (1.3) &
2.33 (0.06) &   6.2 (0.3) & 20.0 (0.7) & 12.8 (0.1) & 10.1 (0.1) \\
\mbox{IRAS04166-R} & $(-20, -35)$ &  \mbox{No~data} & \mbox{No~data}  &
0.27 (0.05) &   13.0 (0.7) &  9.1 (2.4) & - & - \\
\mbox{IRAS04166-B} & $(+20, +35)$  & \mbox{Bad~fit} & \mbox{Bad~fit} &
0.53 (0.04) &  0.0 (0.4) & 10.0 (1.1)  & 12.4 (0.3) & 9.2 (0.1)  \\
\mbox{L1551-B} & $(-255, -255)$  & \mbox{Bad~fit} & \mbox{Bad~fit}  &
 0.36 (0.03) &  -3.2  (0.2) &  5.9 (0.5) & 12.3 (0.2) & 9.2 (0.1) \\
\mbox{L1551-R} & $(+150, +20)$  & 1.3 (0.2) &   13  (1.4) &
 0.58 (0.04) & 7.8  (0.3) &  5.7  (0.6)  & 12.4 (0.1) & 9.7 (0.1) \\
\mbox{L1527-B} & $(+40, +10)$  &  \mbox{No~data} & \mbox{No~data}  &
 0.93 (0.07) &  9.3 (0.8) & 20.3  (2.1) & - & - \\
\mbox{HH1} & $(-30, +55)$ & 2.9 (0.3) &  13  (1.0) &
- & \mbox{No~data} & - & - & - \\
\mbox{HH2} & $(+60, -80)$ &  3.5 (0.5)  &  30 (4.4) &
 2.4 (0.05)  & 9.8  (0.2) &  17.7  (0.4) & 12.9 (0.1) & 9.7 (0.1) \\
\mbox{HH212-B} & $(-15, -35)$  & 0.4 (0.1)  &  37 (13) &
0.76  (0.04) &   1.1 (0.2) & 8.4 (0.5) & 12.3 (0.1) & 9.5 (0.1)  \\
\mbox{HH212-C} & $(0, 0)$  & 2.5 (0.4) &  18 (1.9) &
 2.3 (0.07)  &  2.6 (0.4) &  23.2 (0.8) & 12.6 (0.1) & 9.8 (0.1) \\
\mbox{HH25-C} & $(0, 0)$ &  12.5 (0.9) &  14 (0.7) &
 5.2 (0.06) & 12.4 (0.09) &  16.6  (0.3)  & 13.1 (0.1) & 9.9 (0.1)  \\
\mbox{HH25-R} & $(+36, -57)$ & 3.2 (0.4) &  36 (4.4) &
 4.9 (0.07) & 11.2 (0.04) & 6.9 (0.09) & 13.1 (0.2) & 9.5 (0.1) \\
\mbox{HH111-B} & $(-170, +21)$  &  0.3  (0.1) &    34 (11) &
- & \mbox{Bad~fit}  & - & - & - \\
\mbox{HH111-C} & $(0, 0)$  & 2.0  (0.3) &  12 (1.3) &
0.54 (0.05) &   9.3 (0.6) &  14.6  (1.6) & 12.4 (0.1) & 9.9 (0.1) \\
\mbox{HH46-R} & $(-40, -20)$ &  0.8 (0.2) & 35 (7.3) &
- & \mbox{No~data} & - & - & - \\
\mbox{HH46-B} & $(-10, 0)$ & \mbox{No~data} & \mbox{No~data} &
 1.54 (0.06) &  10.8  (0.4) &  20.0  (1.0)  & - & -    \\
\mbox{BHR71-R} & $(-39, +140)$ &   1.9 (0.2) &  46 (5.5) &
 7.0 (0.08) &  3.0  (0.1) & 16.1 (0.2) & 13.4 (0.5) & 8.8 (0.4) \\
\mbox{BHR71-B} & $(+42, -100)$ & 2.7 (0.3) & 32 (3.4) &
3.4 (0.06) & -6.4  (0.07) & 7.5  (0.2) & 13.6 (0.7) & 8.9 (0.6) \\
\mbox{HH54B}$^{(5)}$ & $(2,4)$  & 8.8 (0.4) &  22 (0.8) &
  10.6 (0.07) & -6.6 (0.05) &  14.5  (0.1) & 13.3 (0.1) & 9.6 (0.1) \\
\mbox{IRAS16293-B} & $(+72, -56)$  & 0.1 (0.2) &  13 (16) &
1.6 (0.05) & 1.7 (0.06) &  3.6   (0.1) & - & - \\
\mbox{IRAS16293-R} & $(+72, +49)$  &  0.8 (0.1) & 38 (6.2)  &
6.2 (0.04) & 8.2 (0.03) &  9.7 (0.07) & 14.1 (0.2) & 7.8 (0.2) \\
\mbox{L483-B} & $(-60, +30)$ &  0.4 (0.2) &  21 (6.3) &
0.70 (0.05) &  1.4  (0.5) &  12.7 (1.0) & 13.6 (0.6) & 8.0 (0.5) \\
\mbox{S68N-B} & $(-12, +24)$ & 6.9 (0.5) & 23 (1.6) &
 4.5 (0.07) &   6.1 (0.1) &  15.5 (0.4) & 13.2 (0.1) & 9.6 (0.1) \\
\mbox{S68N-C} & $(0, 0)$ &  8.9 (1.2) & 31 (3.6) &
10.9 (0.07) &  9.1 (0.05) &  17.6  (0.2) & 13.2 (0.1) & 9.8 (0.1) \\
\mbox{SERSMM1-B} & $(-18, +30)$  & 8.8 (0.8) & 21 (1.9) &
4.5 (0.09) &  10.0  (0.2) &  22.6 (0.9) & 13.2 (0.1) & 9.7 (0.1) \\
\mbox{SERSMM4-B} & $(-60, +30)$  & 0.4 (0.2) & 18 (10) &
 1.9 (0.03) &  5.0 (0.04) &  5.7 (0.1)  & 13.9 (0.2) & 7.6 (0.4) \\
\mbox{SERSMM4-R}$^{(6)}$ & $(+30, -60)$  & 0.4 (0.1) & 44 (21) &
1.9 (0.05) & 11.6 (0.2) &  13.2 (0.6) & 13.1 (0.6) & 8.6 (0.5) \\
\mbox{B335-B} & $(+30, 0)$  & \mbox{No~data} & \mbox{No~data} &
- & \mbox{Bad~fit}  & - & - & - \\
\mbox{N7129FIR2-R} & $(+50, -50)$  &  0.3 (0.2) &  19 (7.1) &
- & \mbox{Bad~fit}  & - & - & - \\
\mbox{CEPE-B} & $(-12, -20)$  & 71.2 (7.7) &  15 (1.1) &
 26.2 (0.2) & -27.4 (0.2) &  44.6 (0.4) & 13.9 (0.1) & 10.0 (0.1) \\
\mbox{CEPE-R} & $(+8, +20)$  & 27.7 (3.4) & 22 (2.1) &
- & \mbox{No~data}  & -  & - & - \\
\hline
\end{tabular}
\begin{list}{}{}
\item[] (1) PACS results from gaussian fits to the radial profiles
of integrated intensity
with rms uncertainty values in parenthesis. The origin of the profile
is the emission centroid and the diameter is
the FWHM of the fitted gaussian (without correction for the $13''$ 
telescope beam);
(2) HIFI results from gaussian fits to the spectra 
with rms uncertainty values in parenthesis. $\Delta V$ represents the FWHM 
of the emission;
(3) Results from the analysis of the combined PACS and HIFI data toward
the emission peak and with a resolution of $13''$, see Sect.~\ref{sec_nt};
(4) Offsets are with respect to the central position in Table~\ref{centers};
(5) Data previously published by \citet{bje11};
(6) Position excluded from sample analysis due to dubious outflow origin.
\end{list}
\end{table*}

\begin{figure*}
\centering
\resizebox{16.7cm}{!}{\includegraphics{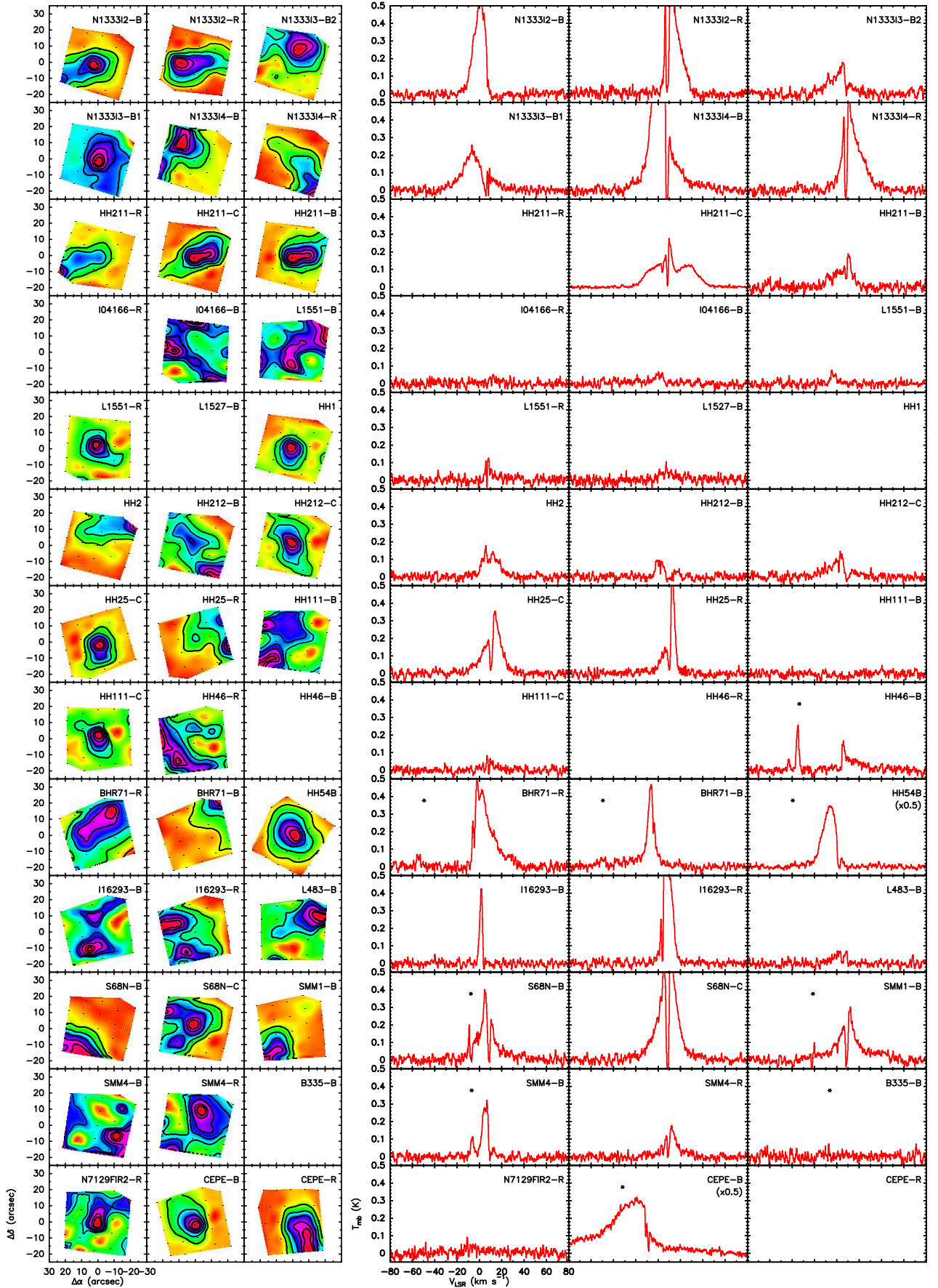}}
\caption{Summary view of all the outflow survey data ordered by 
increasing right ascension, as in Table~\ref{summary}.
{\bf Left panels: } PACS maps
of H$_2$O(2$_{12}$--1$_{01}$) integrated intensity showing contours at 
20, 40, 60, 80, and 90\%
of the peak value (see Table~2 for absolute intensities). 
The points indicate the location of the individual PACS spaxels.
{\bf Right panels: } HIFI spectra
of  H$_2$O(1$_{10}$--1$_{01}$) with fixed intensity and velocity 
scales for easier inter-comparison
(some bright spectra have been scaled down to fit the box). 
Asterisk signs in some spectra indicate the position of the
NH$_3$($JK$=10--00) line coming from the upper sideband of the receiver.
Empty boxes correspond to positions observed with one instrument
but not with the other.
\label{summary-fig}}
\end{figure*}

Fig.~\ref{summary-fig} presents a summary view 
of all the data from the outflow survey.
The left block of panels shows the PACS results in the form
of $5\times 5$ integrated-intensity maps using contours
proportional to the map peak intensity. The right block of panels
presents the HIFI spectra with fixed scale in both velocity and 
intensity. In total, 39 different positions were observed in at least 
one of the two H$_2$O lines, and 32 positions were observed with both 
PACS and HIFI (some positions were dropped during the survey due to
weak emission and time limitations). 

As the figure illustrates, the objects in the sample present a diversity of
 spatial distributions and intensities. The PACS maps show that the
1670~GHz emission  tends to be
spatially concentrated, but that it usually extends over scales 
larger than the $13''$ PACS beam. The emission peaks
do not always coincide with the central position of the map, which
corresponds to our 
expected location for the H$_2$O maximum. An object-by-object inspection 
shows that this mismatch
arises from a combination of
errors in the literature maps used to
prepare the observations and true offsets 
between the
peaks of the H$_2$O emission and the peaks of the molecular emission 
used to choose the PACS map center (usually CO).
The origin of these offsets will be further explored below.

Less clear from the PACS maps due to the use of relative
contours is the large range of intensities covered by the sample. This 
is better appreciated from the HIFI spectra, which cover almost two orders
of magnitude in integrated intensity between the brightest (CEPE-B) 
and weakest (L1551-B)  557~GHz lines. A large
intensity range must be intrinsic to the sample, and cannot arise solely
from errors in predicting the peak position, or from 
beam dilution effects, since sources like L1551-R or HH111-C present
very weak HIFI spectra despite their emission being well centered on the PACS maps.
As we will see below, the large range in integrated intensities
seems to arise from an equivalent large range of H$_2$O column densities
in the sample. This means that although the target selection was biased
toward bright H$_2$O candidates, the sample has still almost two orders of
magnitude of dynamic range, which gives a convenient margin to 
explore different emission conditions and 
optical depth effects in the targets. 

Also noticeable in the HIFI data are the diversity of linewidths
and spectral shapes. As previously noticed by \citet{kri12}
in their observations toward the low-mass YSO themselves, most 557~GHz 
lines present a narrow dip at ambient velocities that likely arises
from self-absorption by low-excitation H$_2$O along the line of
sight (see \citealt{cas12} for a study of ambient H$_2$O emission 
and absorption in dense cores).
An additional narrow feature appears at shifted velocities toward
a number of spectra taken in the first batch of observations, like 
HH46-B. It results from the superposition of NH$_3$(10--00) emission,
coming from the upper sideband of the receiver, and its position 
has been indicated by an asterisk in those spectra where it appears.
Apart from these two narrow features, the HIFI spectra are dominated by
broad wings typical of outflow emission.  

\begin{figure*}
\centering
\resizebox{\hsize}{!}{\includegraphics{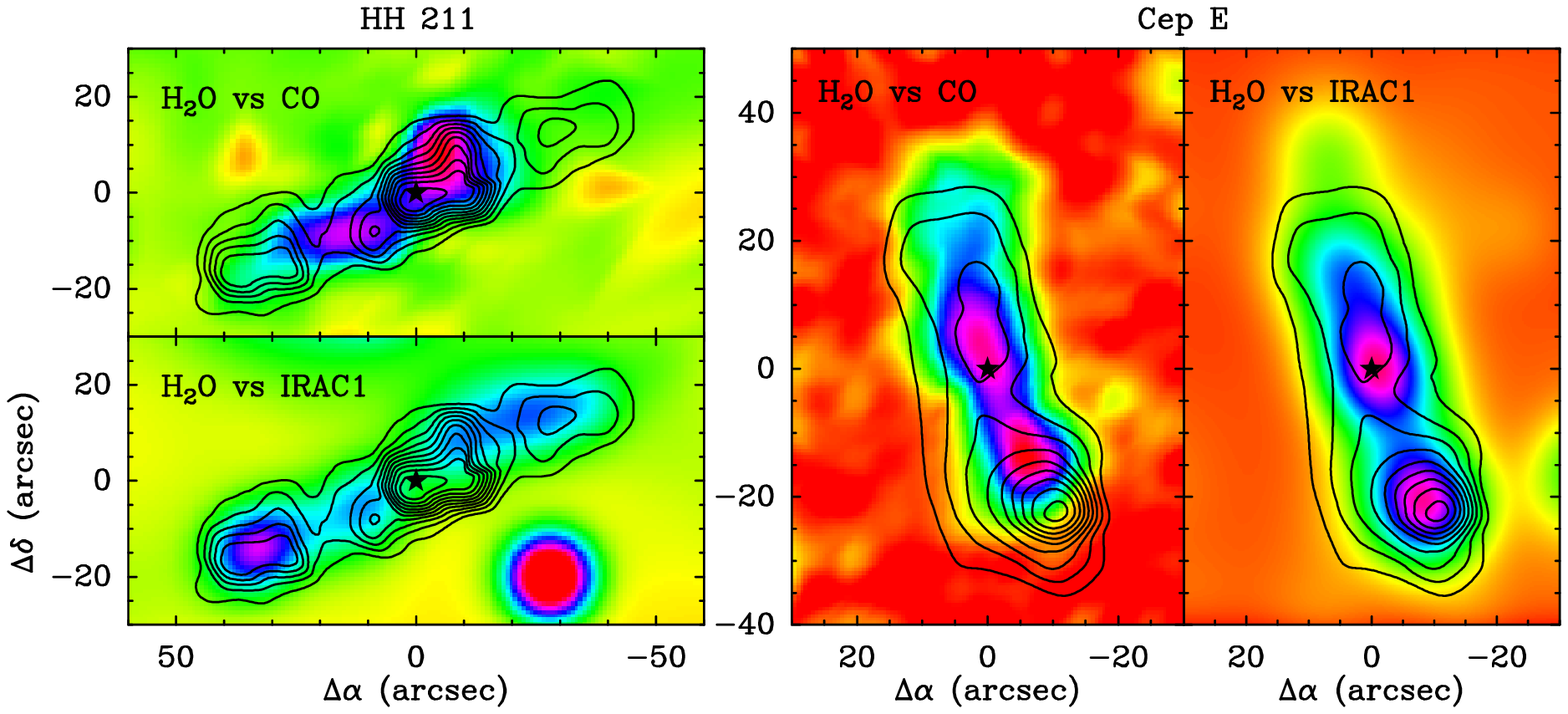}}
\caption{Comparison between H$_2$O(1670 GHz) integrated
intensity (contours) and either CO(2--1) or
H$_2$-dominated IRAC1 images
(color-coded background) for the HH~211 and Cepheus~E outflows.
The IRAC1 image has been convolved to a resolution
of $13''$ to match the resolution of the H$_2$O(1670 GHz) and
CO(2--1) data. For H$_2$O(1670 GHz), 
first contour and contour interval are 1~K~km~s$^{-1}$
for HH211 and 8~K~km~s$^{-1}$ for Cep~E.
The CO map of HH211 uses data presented
in \citet{taf06}, and represents CO(2--1) intensities integrated
in the velocity range $\vert V-V_0 \vert \le 5-20$~km~s$^{-1}$
(ambient cloud velocity $V_0 = 8.6$~km~s$^{-1}$). 
The CO map of Cep~E
represents CO(2--1) intensities integrated
in the range $\vert V-V_0 \vert \le 10-50$~km~s$^{-1}$
(ambient cloud velocity $V_0 = -13$~km~s$^{-1}$), and has been
shifted by $5''$ to the west to correct for a possible pointing
problem suggested by an overlap with the better-registered
interferometer map of \cite{mor01}. 
Note the better agreement of the H$_2$O with the H$_2$-dominated IRAC1
emission than with the CO(2--1) emission.
In all plots, the star symbol indicates the position of the YSO, which is
the origin of the offset values and whose
absolute coordinates are given in Table~1.
The bright circular feature near ($-30''$, $-20''$) in the HH~211
IRAC1 image
corresponds to an unrelated star.
\label{hh211_cepe}}
\end{figure*}

The maps and spectra in Fig.~\ref{summary-fig} also illustrate
the complementarity of the PACS and HIFI observations.
The PACS data lack velocity resolution, but provide information
about the spatial distribution of the H$_2$O emission. They do this with
a relatively high angular resolution of $13''$ over a region
of $47''\times 47''$. The single-pixel HIFI data, on the other hand,
do not provide spatial information, but have a velocity resolution
of 0.6~km~s$^{-1}$. 
The beam size of the HIFI data ($39''$) is similar to the
field of view of the PACS observations, so  
the HIFI velocity-resolved
spectra correspond to an emitting
region approximately
the size of the PACS maps. 
The goal of the analysis presented here is to 
combine the spatial and velocity information provided by 
PACS and HIFI into a self-consistent picture of the H$_2$O emission
from outflow gas. 
As we will see in Sect.~\ref{sec_int_corr}, this approach
is justified by the tight correlation between the intensities 
of the 557 and 1670~GHz lines, which argues strongly for
the two transitions arising from the same volume of gas.
Before combining the PACS and HIFI observations, however,
we study the two sets of observations separately and
characterize the
spatial and velocity properties of the H$_2$O emission.

\section{PACS data: spatial information}

\subsection{Two illustrative outflows: HH~211 and Cepheus E}
\label{sec_hh211_cepe}

Our survey observations were not designed to map the full 
H$_2$O emission
from outflows, which is often extended
and requires dedicated on-the-fly observations.
A separate effort inside the WISH project 
was dedicated to map a selected number of outflows, 
and initial results have already been presented 
\citep{nis10a,bje12,nis13}.
The outflows from the targets HH~211 and Cepheus E, however,
are compact enough to be covered with two or three
PACS fields of view, so our observations provide
full maps of the 1670~GHz H$_2$O emission
in these systems.
Although not as finely sampled as the
dedicated on-the-fly maps,
these small PACS maps can be used to study
the relation between the H$_2$O emission and 
the emission from
other outflow tracers, in particular CO and H$_2$.

Previous observations of HH~211 and Cepheus~E have shown
that the two outflows share a common feature.
Their emission in low-$J$ CO transitions peaks significantly closer to the protostar
than their H$_2$ emission, which is brighter toward the end of the
outflow lobes. This offset between the H$_2$ and low-$J$ CO emitting regions 
is specially noticeable in the maps
of HH~211 by \citet{mac99} (their Fig.~6) and Cepheus~E 
by \citet{mor01} (their Fig.~9). It most likely results from 
the outflows having at least two spatially-separated components 
of different temperature, with 
the H$_2$-emitting gas being 
significantly hotter than the low-$J$ CO-emitting gas
\citep{mor01}. 
This stratification of the outflow emission makes
HH~211 and Cepheus~E  ideal
targets to probe the gas conditions traced by the H$_2$O
emitting gas, and in particular, to distinguish 
between an origin in gas with low excitation (CO-like) and
high excitation (H$_2$-like).

Fig.~\ref{hh211_cepe} presents
a comparison between the emission from H$_2$O and that of
CO and the H$_2$-dominated IRAC1 band
toward HH~211 and Cepheus~E.
In all panels, the contours represent the integrated
intensity of the PACS-observed H$_2$O(1670 GHz) line, 
while the color backgrounds are
the CO(2--1) IRAM 30m emission in the ``H$_2$O vs CO''
panels and the Spitzer/IRAC1 emission in the ``H$_2$O vs IRAC1''
panels.
All the data have similar angular resolution, since
the IRAC1 image has been convolved with a $12''$ 
gaussian, and both the H$_2$O and CO data have 
intrinsic resolutions of $12-13''$.

As can be seen, the  H$_2$O emission from HH~211
presents three separate peaks, one toward the YSO
and one toward the end of each outflow lobe.
The CO emission, on the other hand, has a bipolar
distribution that consists of two 
peaks approximately located half way between the
central source and the outer H$_2$O peaks.
While not completely anti-correlated, the H$_2$O and CO
emissions clearly do not match and their peaks seem
to avoid each other.
In contrast with CO (and in agreement with NIR H$_2$ images),
the H$_2$-dominated IRAC1 emission peaks further 
from the YSO and matches 
better the H$_2$O emission at the end of the two lobes,
especially toward the brightest south-east end of the
outflow.
No IRAC emission is seen toward the central H$_2$O peak, but this
may result from strong extinction, since even
the protostellar continuum is invisible in the IRAC bands.

The better match between the H$_2$O and IRAC1 emissions
is also noticeable in Cepheus~E (Fig.~\ref{hh211_cepe}
right panels). As in HH~211, the CO emission from the 
southern outflow lobe lies closer to the YSO than the 
H$_2$O emission, while the IRAC1 emission matches
well the bright southern H$_2$O peak.
Less clear is the comparison toward the northern lobe, 
since all emissions drop gradually away from the YSO
(the IRAC emission toward the YSO is likely
contaminated by protostellar continuum, see \citealt{nor04}).
In any case, the maps in Fig.~\ref{hh211_cepe} show 
that the H$_2$O emission from Cepheus~E is, like
in HH~211, more H$_2$-like than CO-like.

\begin{figure}
\centering
\resizebox{\hsize}{!}{\includegraphics{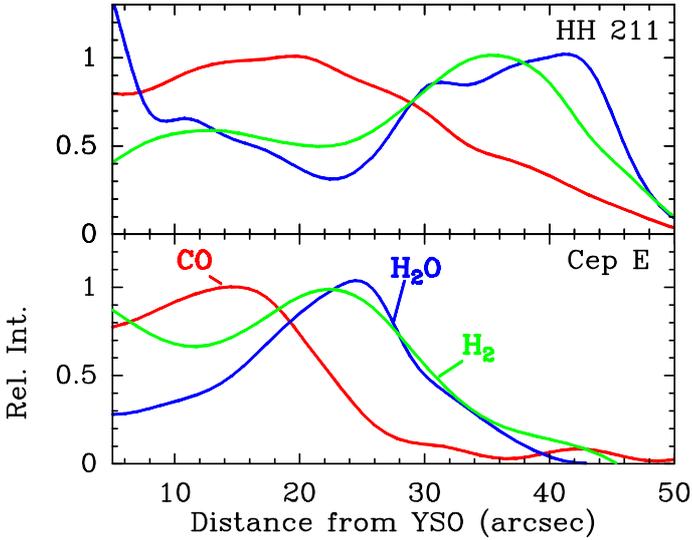}}
\caption{Spatial profiles of the emission from Fig.~\ref{hh211_cepe}
along the eastern lobe of HH~211 (top) and
southern lobe of Cepheus~E (bottom).
The blue lines represent the
1670~GHz H$_2$O emission, the
green lines represent H$_2$-dominated IRAC1 emission, and the
red lines correspond to CO(2--1). All data have a similar resolution
of approximately $13''$.
\label{cuts}}
\end{figure}

A more quantitative comparison between the
H$_2$O, CO, and H$_2$ emissions is presented in Fig.~\ref{cuts}
using intensity cuts along the outflow
axes for both the eastern lobe of HH~211 and the southern lobe of 
Cepheus~E.
These two lobes present 
the brightest H$_2$O and H$_2$ intensities  \citep{mac99,mor01}, and are therefore
the best regions for a comparison between the different outflow tracers. 
As can be seen, the  H$_2$O and H$_2$ emissions 
(blue and green lines) peak approximately at the same distance from the YSO
and have similar widths, while the CO emission
(red line) peaks closer to the YSO by $15''$ in HH~211 
and $10''$ in Cepheus~E. 
The close match between the H$_2$O and H$_2$ spatial profiles
indicates that the gas conditions responsible for the two
emissions must be rather similar, while they must 
differ significantly
from the conditions of the gas responsible for the CO(2--1) emission.
This is a first indication that the H$_2$O-emitting gas in the
outflow lobes has
a higher excitation than the low-$J$ CO-emitting gas
commonly associated with outflow material.

\subsection{A general correlation between the H$_2$O and
H$_2$ emissions}
\label{sec_h2o_h2}

The spatial correlation between the H$_2$O  and H$_2$ emissions 
is not limited to the HH211 and Cep~E outflows just studied,
but seems to extend to the whole sample.
A one-by-one comparison between the 
PACS maps of Fig.~\ref{summary-fig} and equivalent IRAC images
from the Spitzer archive shows that in most cases the H$_2$O emission
matches spatially that of H$_2$, even when 
the H$_2$ and the low-$J$ CO emissions differ in their distribution
(like seen in HH 211 and Cep E).
The PACS H$_2$O maps are therefore systematically ``H$_2$-like''
both in peak location and spatial extent, 
and this suggests that the conditions
of the gas responsible for the H$_2$O emission 
are similar to those of the H$_2$-emitting gas.

The similar spatial distribution
of the H$_2$O and H$_2$ emissions
was not recognized at the time of target
selection (circa 2007), and this explains why 
a number of PACS maps in  Fig.~\ref{summary-fig}
appear offset or even miss the H$_2$O peak.
Target selection in our survey 
was mainly guided by low-$J$ CO maps, 
so most PACS centers were chosen to 
coincide with the peak of this relatively low excitation emission.
L483 provides a good illustration of this issue, since
in this outflow the H$_2$ peak is known to lie more than $20''$
to the west of the CO peak
\citep{ful95,taf00}.
As Fig.~\ref{summary-fig} shows, our CO-centered 
PACS map misses a significant part of the H$_2$O emission,
which extends to the west of our chosen field of view.
Although unfortunate, the sometimes dramatic effect of our
shifted target selection has helped to highlight the H$_2$-like
nature of the  H$_2$O emission. It also has made our 
H$_2$O survey cover not only the bright emission peaks but 
the more extended component.

A notable exception to the good match between PACS and IRAC
images is the NGC~1333-I2 outflow. The PACS maps of this source present two
bright H$_2$O peaks that coincide 
with the CO/CH$_3$OH/SiO outflow maxima east and west of the YSO
\citep{san94,bac98,van98,joe04},
while no H$_2$ emission from either H$_2$O peak can be discerned in the 
IRAC images.
Although this may indicate an anomalous 
behavior of the NGC~1333-I2 outflow, it  more 
likely results from high extinction inside the 
NGC~1333 star-forming dense core.
This interpretation is supported by 
the scarcity of background stars seen by IRAC 
and by the recent observations at
longer wavelengths by \citet{mar09}.
These authors found
a bright H$_2$ $S(1)$ 17$\mu$m emission peak toward 
the eastern lobe of NGC~1333-I2 with similar shape and size
to the H$_2$O peak seen in the PACS map.
Such detection of $S(1)$ emission
indicates that at least the eastern lobe of the NGC~1333-I2
outflow
is associated with a significant amount of excited 
H$_2$, and that if this emission is not seen in the IRAC images,
it is likely due to an extreme case
of extinction similar to that occurring
at center of the HH211 outflow.
Unfortunately, \citet{mar09}
did not cover the western lobe of the outflow 
in their map,
so the status of this position remains uncertain.

\begin{figure}
\centering
\resizebox{7cm}{!}{\includegraphics{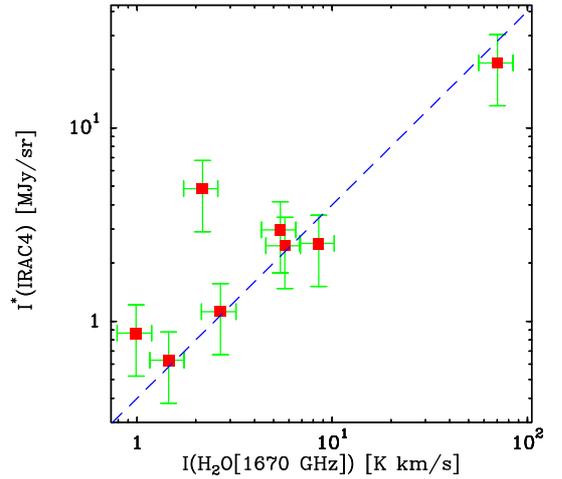}}
\caption{Comparison between extinction-corrected 
IRAC4 intensities and H$_2$O(1670 GHz) integrated intensities for the sources of 
Table~\ref{irac4}. The formal error bars assume a 40\% uncertainty in the IRAC4
intensities (due to the accumulated uncertainty of background subtraction
and extinction correction) and a 20\% uncertainty in the PACS intensities.
The dashed line
is a linear correlation that, according to the analysis of section~\ref{sec_abu},
corresponds to a constant H$_2$O abundance value of $3 \times 10^{-7}$.
\label{irac4_pacs}}
\end{figure}

The correlation between the
H$_2$ and H$_2$O emissions 
is not limited to morphology, but 
involves line intensities.
Comparing the intensities in the PACS and IRAC images,
however,
is not a straightforward operation,
since the IRAC intensities represent more than
just H$_2$ emission. They contain possible contributions from 
continuum emission from YSOs and unrelated objects together 
with diffuse background radiation from the cloud (plus the already
mentioned non-negligible dust extinction
in dense regions).
To minimize these effects, we limit our PACS-IRAC 
comparison to the peak values of
positions where the IRAC emission can be reasonably
expected to have uncontaminated H$_2$ origin 
and to be
associated with the H$_2$O emission seen with PACS. 
We have done this by selecting
the sources whose well-defined PACS maximum 
is offset more than
$10''$ from the YSO position (to avoid protostellar
continuum contribution in the IRAC images). 
For these sources, we have convolved
the IRAC images with a gaussian to simulate the $13''$ resolution
of the PACS observation, and used this convolved image
together with the PACS map to estimate the H$_2$ and H$_2$O intensities
at the peak. In order to subtract the extended emission contribution
(important in the IRAC images)
we have measured in each image the intensities at three different positions:
the H$_2$O(1670 GHz) peak and two off-peak positions that seem unaffected 
by protostellar or background contamination. The average intensity
of these off-center positions is used to
estimate a background contribution, which is then subtracted
from the peak intensity.
A further correction of the IRAC intensity for dust extinction is
made using literature values of A$_V$ extrapolated to the
IRAC wavelengths, assuming A$_K$/A$_V$ = 0.112 \citep{rie85}
and the A$_\lambda$/A$_K$ ratios recommended by \citet{ind05}.

\begin{table}
\caption[]{H$_2$O(1670 GHz)-IRAC4 correlation}
\label{irac4}
\centering
\begin{tabular}{l c c c c}
\hline
\noalign{\smallskip}
\mbox{Source} & \mbox{$I$[H$_2$O(1670 GHz)]}  &
\mbox{$I$[IRAC4]} & \mbox{A$_V$} & \mbox{A$_V$ Ref.}  \\
& \mbox{(K~km~s$^{-1}$)} & \mbox{(MJy sr$^{-1}$)}
& \mbox{(mag)} \\
\noalign{\smallskip}
\hline
\noalign{\smallskip}
\mbox{N1333I3-B2} & 2.2 & 3.4  & 9 & (1) \\
\mbox{N1333I3-B1} & 5.4  & 2.1 & 9 & (1) \\
\mbox{HH211-B} & 5.7 & 1.7 & 8 & (2) \\
\mbox{HH1} & 2.7 & 1.1 & 1.5 & (1) \\
\mbox{HH46-R} & 1.0 & 0.6 & 8 & (3) \\
\mbox{BHR71-R} & 1.5 & 0.6 & 1 & (4) \\
\mbox{CEPE-B} & 70.3  & 12.5 & 12.5 & (5) \\
\mbox{HH54} & 8.5  & 2.3 & 2 & (4) \\
\hline
\end{tabular}
\begin{list}{}{}
\item[] (1) \citet{gre96}; (2) \citet{dio10a}; (3) \citet{fer00};
(4) \cite{car06}; (5) \citet{smi03}
\end{list}
\end{table}

\begin{figure*}
\centering
\resizebox{\hsize}{!}{\includegraphics{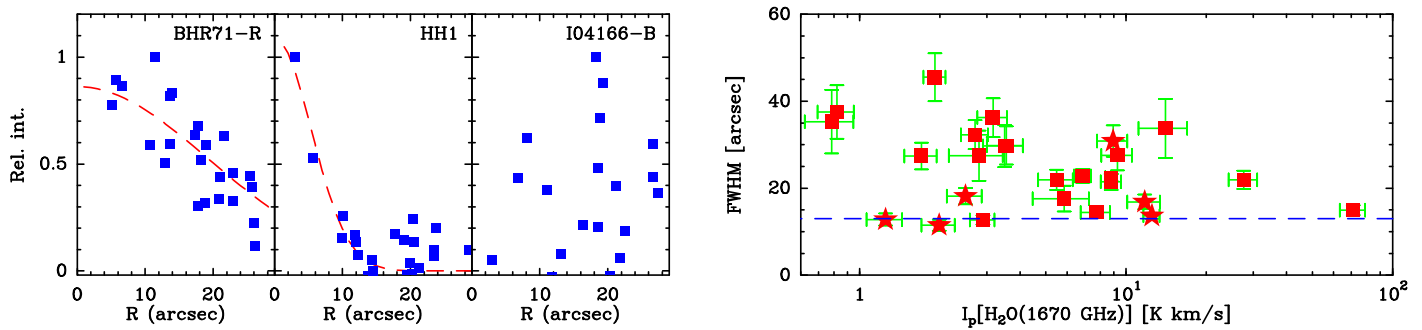}}
\caption{{\bf Left: } sample of intensity radial profiles
illustrating the different cases encountered in the analysis
of the 1670~GHz line PACS data:
outflow with extended emission (BHR71-R),
outflow with compact emission (HH1), and outflow
with emission that is too weak to allow a meaningful fit (I04166-B).
Blue dots are PACS data and red dashed lines
are gaussian fits.
{\bf Right: } comparison between the size (uncorrected
for the telescope beam) and intensity of the H$_2$O-emitting
region as determined from the gaussian fits illustrated
in the left panel. The star symbols
indicate data from maps centered at a YSO position and the horizontal
dotted line indicates the telescope FWHM at 1670~GHz ($13''$).
\label{pacs_sizes}}
\end{figure*}

Fig.~\ref{irac4_pacs} compares the H$_2$O(1670 GHz) and extinction-corrected 
IRAC4 intensities
for the objects that passed our selection criteria
(see Table~\ref{irac4} for numerical values and notes).
Although there is a good correlation between the H$_2$O and IRAC 
intensities at all IRAC bands,
we focus on IRAC4
because its passband includes the                         
lowest H$_2$ rotational transitions observable
by IRAC ($S(4)$ and $S(5)$),
and is therefore less sensitive to the 
small fraction of very hot gas that dominates
IRAC1 and IRAC2 observations \citep{neu08}.
As can be seen in Fig.~\ref{irac4_pacs}, 
there is a reasonable correlation between H$_2$O(1670 GHz) and
extinction-corrected
IRAC4 intensities that covers almost two orders of magnitude
in range and has a Pearson r-coefficient of 0.98.
We approximate this correlation with the
simple expression
$$I^*[\mathrm{IRAC4}]\; (\mathrm{MJy~sr}^{-1})  = 
0.4\; I[\mathrm{H}_2\mathrm{O(1670)}] \; (\mathrm{K~km~s}^{-1}),$$
where $I^*$[IRAC4] is the extinction-corrected 
IRAC4 intensity.
This correlation is
indicated by the dashed line in the figure,
and is closely followed by the objects
with best-defined emission peaks in
both H$_2$O and IRAC maps: HH211-B, HH54, and CEPE-B.
The two objects that lie significantly above
the dashed line in Fig.~\ref{irac4_pacs}
are N1333I3-B2 and HH46-R, which
have poorly defined IRAC4 peaks
whose intensity
may have been overestimated.

The correlation between H$_2$O and IRAC4 intensities
has a number of implications. 
It supports the relation between the H$_2$O and H$_2$ 
emission initially inferred from the similarity of their
spatial distributions, and shows that 
outflows located in different clouds and
powered by sources of different luminosity
share a common ratio between H$_2$O and H$_2$ 
intensities.
Since the H$_2$ emission is generally optically thin 
and approximately proportional to the 
H$_2$ column density \citep{neu08}, the H$_2$-H$_2$O correlation 
suggests that the H$_2$O emission
must have similar properties.
If so (and the LVG analysis of Sect.\ref{sec_lvg} confirms it),
the correlation implies that the emitting gas H$_2$O abundance must
be close to constant over the sample. 
Calculating the exact value of this abundance requires
determining the excitation conditions of H$_2$O, and for
this reason, we defer the discussion to
Sect.~\ref{sec_abu}, where we analyze the combination
of the PACS and HIFI data.

\subsection{Angular size of the emitting region}
\label{sec_size}

The PACS maps of Fig.~\ref{summary-fig} illustrate the
variety of sizes and distributions
seen in the H$_2$O emission. Despite this variety, 
a common feature stands out:
most maps are compact and present
well-defined peaks surrounded by 
more diffuse emission.
Such relatively small emission sizes 
testify to the rather special conditions needed
to produce the H$_2$O emission, and
raise the possibility that 
beam dilution has affected the appearance
of the maps and has decreased artificially the
observed intensities.
To asses this possibility, we
quantify the size of the emitting region
in the PACS maps.

Given the large variety of sizes and shapes seen in
the maps, any attempt to condense all the 
spatial information into a single
``size'' parameter
is necessarily an approximation.
Our goal in this section, however, 
is not to characterize in detail any 
of the individual objects,
but to derive a statistical estimate of the 
water-emission size to assess from it
the effect of the PACS finite angular resolution. 
For this reason, we
have chosen the simple approach of
fitting a gaussian to the radial 
profile of emission in each of our PACS images.
To do this, we
have first determined the emission centroid 
using all positions having an intensity 
at least half the value of the map peak
(to minimize noise effects).
Using this centroid, we have created a
radial profile of emission, and we have fitted it
with a one-dimensional gaussian using a standard
least-squares routine (part of the GILDAS analysis 
package). A sample of radial profiles and their fits 
are shown in the left panels of Fig.~\ref{pacs_sizes},
and the resulting estimates of the emission size 
and peak intensity are presented
in Table~\ref{summary}.

The right panel of Fig.~\ref{pacs_sizes} presents our estimated
H$_2$O emission sizes as a function of
peak intensity for all
26 sources in the sample whose gaussian fit
parameters were determined with S/N larger than 3. 
The only noticeable trend seen in the plot is a
generally smaller 
size for sources that are centered on
a YSO position, and which are represented in the
figure by star symbols. Several of these sources
present values close to the $13''$ PACS FWHM 
(horizontal dotted line), and are therefore
consistent with being unresolved.
Apart from this trend, no clear correlation between size 
and intensity can be seen in the plot, and most
points seem to be randomly scattered between $13''$
and $40''$. The two brightest positions in
the diagram correspond to
the Cepheus~E outflow, and their smaller size may be 
partly 
enhanced by the larger distance to this source
compared to the others in the sample
($\approx 700$~pc
compared to $\approx 300$~pc of most other sources).

Using the 26 points shown in Fig.~\ref{pacs_sizes}, we estimate a
mean FWHM size for the H$_2$O-emitting region
of $24.5''$, with an rms of $9''$. 
This rms value most likely reflects a scatter in the
true sizes of the sources, as illustrated in Fig.~\ref{pacs_sizes},
and is not simply a result of deviations from gaussian shape
in the radial profiles (although this effect is not negligible).
Deconvolving each fitted FWHM  by subtracting in quadrature a 
$13''$ gaussian (and assuming zero size if the fit
value was smaller than $13''$), we estimate a typical 
intrinsic mean source size of $19.4''$ with an rms of $12''$.
We thus conclude that 
apart from a handful of point-like
sources, mostly associated with YSO positions,
the H$_2$O emission in the outflow gas 
is slightly but significantly 
extended compared to the PACS $13''$ beam size. 
As a result, dilution factor corrections to the PACS
intensities are not expected to
be significant (80\% of positions not coincident
with a YSO require dilution correction less than 2).
Of course, a finite size of the emitting region 
does not imply the absence of unresolved
features in the emission. It means
that any compact component must be
accompanied by extended emission, and that
the integrated intensity inside the PACS map
has a larger contribution from the extended
emission than from the unresolved feature.

\section{HIFI data: velocity information}

The $39''$-resolution HIFI observations of the 557~GHz 
H$_2$O line complement the  1670~GHz PACS data
by providing velocity information over a region comparable
to the PACS field of view.
In this section we analyze the
HIFI observations of our outflow sample
with emphasis on their
statistical properties, and in particular on the information
they provide about the velocity properties
of the H$_2$O-emitting gas. 

As Fig.~\ref{summary-fig} shows, there is a 
large range of line shapes and outflow velocities
in the HIFI spectra, 
indicative of the wide variety of outflows present our sample.
The fastest H$_2$O emission corresponds to
the blue lobe of the Cepheus~E outflow, with a maximum velocity of
100~km~s$^{-1}$ with respect to the ambient cloud. Next are
the red lobes of the BHR71 and NGC1333-I4
outflows, which have values close to or higher than 40~km~s$^{-1}$.
Such high velocities are comparable to those found by
\citet{kri11} and  \citet{kri12} toward the position of
the protostellar sources
(although they are significantly lower than
those of some H$_2$O masers
in high-mass star-forming
regions, e.g., \citealt{mor76}).
Together, they
attest the resilience of the H$_2$O molecule,
and its likely formation
in fast post-shock gas.

\subsection{Parameterizing the HIFI spectra}

To compare the properties of the H$_2$O emitting gas
in the different outflows of our sample we need to condense
the variety of observed line shapes 
into a small set of parameters.
A simple-minded but effective approach is to
fit gaussian profiles to the spectra,
and to use the fit-derived parameters as first order
estimates of the emission properties.
To carry out the fits, we first
mask all channels in each spectrum that show
evidence for contamination by NH$_3$
or that display hints of
self absorption by unrelated cold
ambient gas. Then, we fit the blanked spectrum 
with a gaussian profile, and we inspect the result visually
to ensure that the fit is meaningful.

Although a symmetric gaussian profile
is not 
the ideal fit to an outflow
spectrum, the two main parameters
of the fit, the peak intensity and
the line width, provide reasonable estimates of the
intensity and velocity spread of the outflow
H$_2$O emission. A test comparison between the
integrated intensity under the gaussian fit and a
more standard estimate based on the integral
of the spectrum using the extreme outflow velocities
reveals an agreement better than 10\%,
which is below the calibration uncertainty of the HIFI data
(Sect.~2.2). We thus conclude that the gaussian fit
returns a meaningful, zeroth-order characterization of the H$_2$O emission.
The results of this fit are summarized in Table~\ref{summary}.

\subsection{Line shapes}
\label{sec_lineshape}

\begin{figure}
\centering
\resizebox{\hsize}{!}{\includegraphics{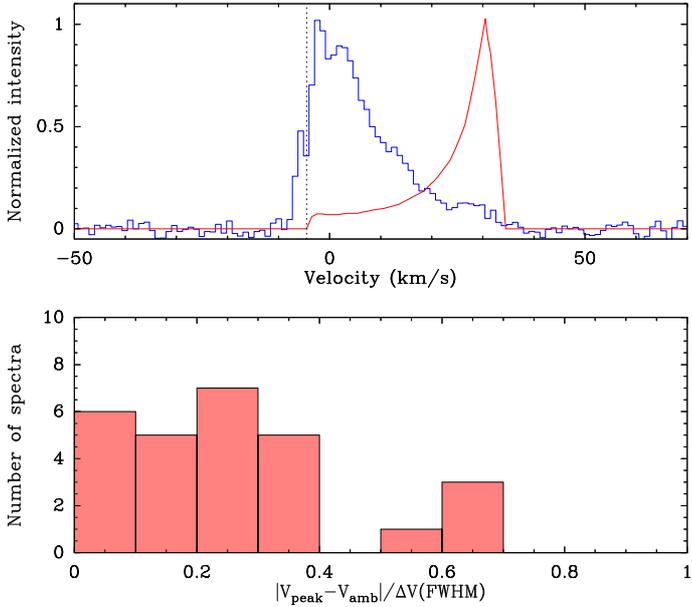}}
\caption{{\bf Top:} comparison between a representative
557~GHz line profile (BHR71-R, blue histogram and maximum velocity
of 40~km~s$^{-1}$) and
a prediction from the planar shock model of 
\citet{flo10} (red line, shock velocity of 40~km~s$^{-1}$
and $n_H = 2\times 10^5$~cm$^{-3}$). The vertical dotted line
indicates the ambient cloud speed. Note the very different shapes.
{\bf Bottom:} Histogram of the ``outflow peak velocity shift''
determined from the 557~GHz spectra
shown in Fig.~\ref{summary-fig}. The shift corresponding to 
the planar shock model in the top panel is 4.2 and lies outside
the range of observed values.
\label{vel_displ}}
\end{figure}

Although the presence of high-velocity emission
is the most noticeable feature of the HIFI spectra, the 
wing shape of the lines implies
that at each outflow position, 
most of the H$_2$O-emitting gas has
relatively low speeds.
This was already seen in the 
spectra of Fig.~\ref{summary-fig}, 
and is illustrated in Fig.~\ref{vel_displ} with 
the HIFI spectrum toward BHR71-R (blue histogram).
The observed wing-like profiles
imply that at each outflow
position, the amount of gas decreases 
systematically with velocity, and therefore,
that the H$_2$O emission is dominated by the slowest 
gas in the outflow.
Of course, wing-like
profiles are typical of 
outflow tracers such as CO, but
in those species, the outflow contribution
can be potentially contaminated with
emission from the ambient cloud.
The selective nature of H$_2$O 
guarantees that the emission arises from warm 
shocked outflow material (Sect.~6),
and indicates that 
predominance of low-velocity material
must be an intrinsic characteristic of the
shock-accelerated gas.

The observed wing-dominated  H$_2$O  profiles 
differ significantly from those predicted by
the planar-shock models commonly used to interpret
H$_2$O emission.
\citet{flo10}, for example, have recently modeled
molecular lines observable
with the Herschel Space Observatory
and generated synthetic spectra of the 557~GHz
H$_2$O line that should be directly comparable with
our observations (see their Fig.~8).
As illustrated by the red line 
in the top panel of Fig.~\ref{vel_displ},
these planar-shock model
spectra present a narrow component that is
approximately centered 
at the shock velocity and that has a weak wing 
toward the ambient cloud speed.
Such spike-like line profile 
is almost a 
mirror image of the observed line profiles
and therefore seems inconsistent with our observations.

To quantify the discrepancy between our observations and
the predicted model spectra, we have
defined a simple parameter that we will refer to as the
``outflow peak velocity shift.'' 
This parameter quantifies the 
effect of the outflow in  shifting the 
velocity of the emission peak in the spectrum, and is
equal to the difference in velocity between the H$_2$O
peak and the ambient cloud 
(determined from N$_2$H$^+$ or NH$_3$ data
and given
in Table.~\ref{centers}) divided by the
FWHM of the H$_2$O spectrum.
As illustrated by the top panel of Fig.~\ref{vel_displ},
line profiles dominated by wing emission are
expected to have outflow peak velocity shifts lower than
unity, while spike-dominated spectra are
expected to have shifts significantly larger than
1 (the planar-shock model spectrum in the figure has a shift
of 4.2).

The bottom panel of Fig.~\ref{vel_displ}
shows a histogram of the outflow peak velocity shifts for the
27 sources in our sample that have peak emission larger than 0.1~K
and a meaningful gaussian fit.
As expected from the wing-like type of profiles,
the histogram is dominated by peak velocity shifts 
close to zero, and no shift exceeds unity.
Many of the small velocity shifts are in fact upper limits, since
the self-absorption 
feature at ambient speeds tends to artificially move
the H$_2$O emission peak away from the ambient cloud velocity.
Even without correcting for this effect,
the outflow peak velocity shifts in our sample
are extremely small, and have   
a mean value of 0.26 with an rms of 0.19. 
For comparison, we have estimated outflow velocity shifts 
for the model spectra of \citet{flo10} using the examples 
shown in their Fig.~8. The values lie
in the range 2-6 with a mean of approximately 
4. Such large values exceed our observed mean shift 
by more than one order of magnitude, and make the
models lie outside the range of velocity shifts covered 
by the histogram in Fig.~\ref{vel_displ}.

The large discrepancy between observed and model-predicted 
H$_2$O line shapes is a strong indication that
the plane-parallel shock approximation used by the models
is not a good representation of the outflow velocity field.
Because of the 1D geometry, the gas in a plane parallel shock
cannot escape the compression and piles up at a single
velocity downstream, producing a 
spike-like feature in the spectrum\footnote{The 
spike-like ``extremely high velocity''
(EHV) component seen in a small group of outflows 
most likely results, not from a planar shock, but 
from a protostellar jet traveling almost ballistically 
along the outflow axis \citep{san09}.
H$_2$O emission from this EHV component has been reported
by \citet{kri11}}.
To avoid this spike and produce the multiplicity of 
velocities characteristic of a wing-like profile,
a more complex velocity field is required.
Numerical simulations show that
bow-shock acceleration by a precessing or pulsating jet
can produce an
increase in the range of velocities of
the outflow swept-up gas (e.g., \citealt{smi97,dow03}).
Models of wide-angle winds 
interacting with infalling envelopes seem to also 
produce a significant mix of velocities \citep{cun05}, 
although more detailed work is needed to explore 
the kinematics of this family of solutions.
While clearly more complex than planar shocks, 
these 2D geometries (or alternative, e.g., \citealt{bje11})
seem necessary to produce the realistic line profiles needed
to properly compare models of shock chemistry with outflow
observations.

\section{Comparison between the 557~GHz, 1670~GHz, and CO(2--1) emissions}
\label{compare}

\subsection{Intensity correlations}
\label{sec_int_corr}

In Sect.~\ref{sec_h2o_h2} we saw that the 1670~GHz line traces an outflow 
component similar
to that responsible for the H$_2$ emission, and therefore,
hotter than the gas emitting  CO(2--1).
Now we investigate whether the 557~GHz line traces the same gas component,
and therefore arises from hot outflow gas, 
or traces the colder outflow material responsible for the CO(2--1) emission.
To do this, we first need to convolve
both the CO(2--1) and 1670~GHz data
to the $39''$ angular resolution of the HIFI 557~GHz observations
so they can be properly compared.
Our on-the-fly IRAM 30m CO(2--1) data cover a region $80''\times 80''$
with Nyquist sampling, so their convolution to $39''$ is straightforward.
The PACS 1670~GHz data cover a region
$47''\times 47''$ with an array of 25 spectra, and although the coverage
is not Nyquist sampled, the data provides enough information
to simulate an observation with $39''$ resolution.
Thus, from now on, our comparisons will use line data
that have an equivalent resolution of $39''$.

\begin{figure*}
\centering
\resizebox{\hsize}{!}{\includegraphics{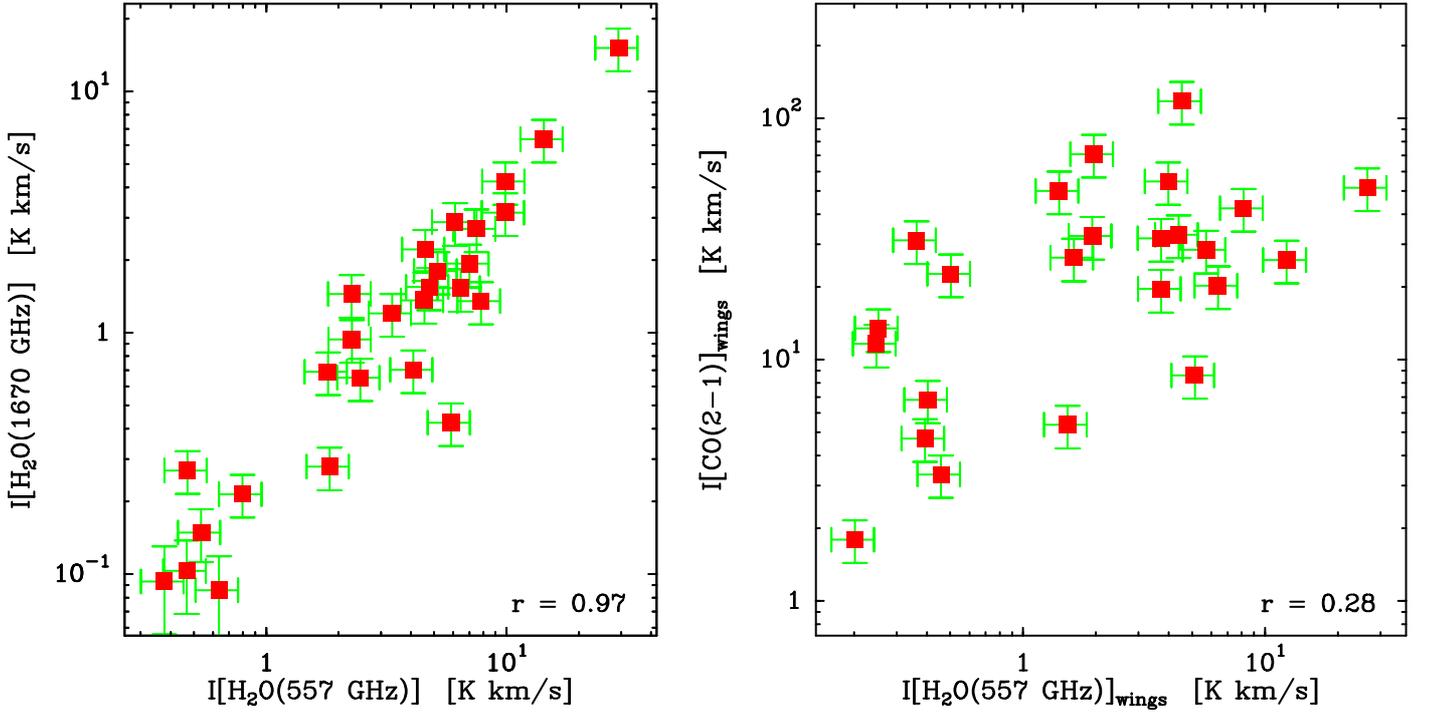}}
\caption{Integrated intensity of the H$_2$O(557~GHz) line vs
H$_2$O(1670~GHz) ({\bf left}) and CO(2--1) ({\bf right}).
Note the tighter correlation in the 557-1670~GHz panel that
suggests the 557~GHz emission arises from 
the same gas that emits the 1670~GHz line.
The Pearson-r coefficients of each correlation are 
indicated in the  bottom-right corner of the
panel.
\label{correl}}
\end{figure*}

Fig.~\ref{correl} presents a comparison between the integrated intensities 
of the 557~GHz line and those of the 1670~GHz and CO(2--1) lines for 
all objects in the outflow sample for which the
required data are available.
The left panel compares the intensities of the 557~GHz and 1670~GHz
lines as derived from integrating their intensity over all velocities.
Using integrated intensities for the 1670~GHz line
is unavoidable due to the lack of effective
velocity resolution in the PACS data.
For consistency, we have integrated  the 557~GHz line profile
over all velocities for which the emission was detected,
simulating a velocity-unresolved observation.
As can be seen, there is a tight correlation between the intensities
of the 557 and 1670~GHz lines over the two orders of magnitude covered
by our data. 
The scatter of points with respect to a linear fit
(in log scale) is low, and the
 Pearson's r coefficient of the dataset is 0.97. This
implies that the
correlation between the intensities of the 557 and 1670~GHz lines
is statistically significant.

In contrast with the correlation between the two H$_2$O transitions, 
the right panel of Fig.~\ref{correl} shows that the
557~GHz and CO(2--1) lines behave almost
independent of each other.
This right panel presents the
557~GHz and CO(2--1) line intensities with
the same logarithmic scale as in the plot of 
the 557 and 1670~GHz intensities, so the two scatter plots in 
the figure can be compared directly.
To avoid contamination from the bright ambient cloud in the 
CO(2--1) emission, the intensities shown in the 557~GHz-CO(2--1)
scatter plot exclude the contribution from the 
central 6~km~s$^{-1}$, which according to an inspection
of the spectra is the maximum range of the ambient emission
in the objects of our sample.
Because of this velocity exclusion, the
right panel of Fig.~\ref{correl} compares intensities in
the outflow regime only, and is independent of contributions from
ambient cloud emission, absorption, or even contamination
from the reference position. 
As can be seen in Fig.~\ref{correl},
the scatter in the 557~GHz-CO(2--1) plot is
higher than in the 557-1670 GHz plot to the left,
and the Pearson r-coefficient is only  0.28. This indicates
that any correlation between the 557~GHz and CO(2--1) 
intensities has only a very low statistical significance.
(Including the contribution from the ambient 
cloud regime further degrades the correlation and 
decreases the r-coefficient.)

The plots in Fig.~\ref{correl} help
answer the question of whether the gas responsible for the
557~GHz line emission resembles the 1670~GHz-emitting gas 
or the one producing CO(2--1).
As can be seen, the
557~GHz intensity is significantly more correlated with the 
1670~GHz intensity than with CO(2--1), 
and this stronger correlation suggests that the 
557~GHz-emitting gas is more closely connected to the 
1670~GHz-emitting gas than to the gas
responsible for the CO(2--1) line.
The correlation between the 557 and 1670~GHz intensities
is in fact so tight and uniform over the two orders of magnitude in
intensity covered by our sample that it seems
unavoidable to conclude that the 
two H$_2$O lines arise from the same volume
of gas.

A common origin of the 557 and 1670~GHz H$_2$O emissions 
also helps explain 
the weak correlation between the
557~GHz and CO(2--1) intensities. In Sect.~\ref{sec_h2o_h2}, we saw that
the 1670~GHz and CO(2--1) emissions are often spatially
offset, and that they likely arise from different volumes of outflow gas.
Our finding now that 
the 557~GHz emission arises from the 1670~GHz-emitting gas implies that
the 557~GHz emission should also be
spatially offset from the CO(2--1) emission, even if the effect cannot be
directly resolved with the low angular resolution of the HIFI observations.
This different physical origin of the 557~GHz and CO(2--1) emissions 
seems the likely cause of the only weak correlation 
between the 557~GHz and CO(2--1) intensities in 
the scatter plot of Fig.~\ref{correl}.

\subsection{Spectral profiles}
\label{sec_h2o_co}

\begin{figure}
\centering
\resizebox{\hsize}{!}{\includegraphics{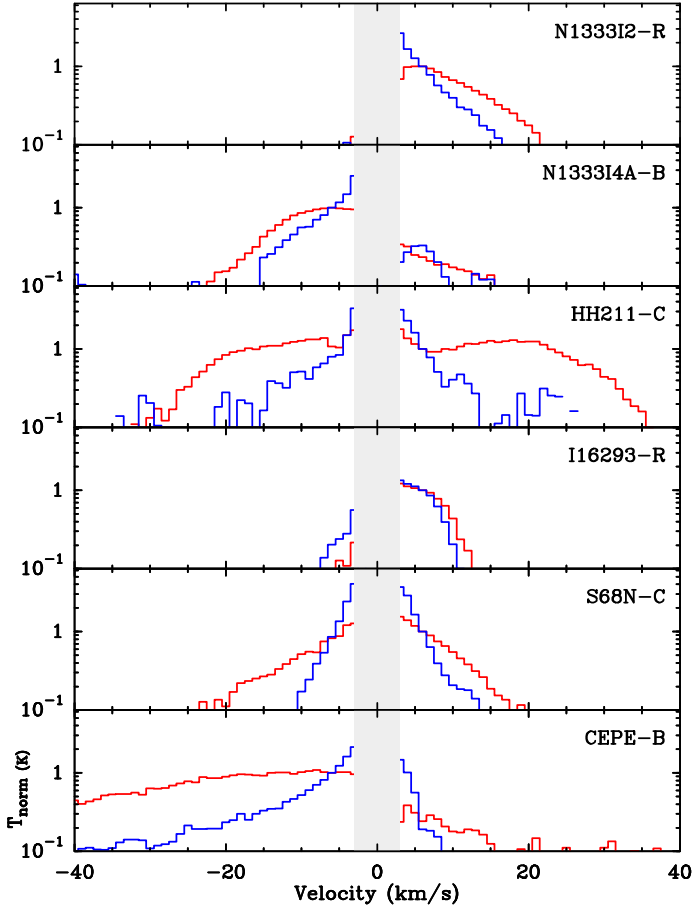}}
\caption{Comparison in log scale between spectra of H$_2$O(557~GHz)
(red) and CO(2--1) (blue) for the six brightest outflows
of our sample illustrating the systematicly flatter 
wings in the H$_2$O line.
For ease of comparison, all spectra have been re-centered 
at zero velocity, and the intensities have been 
normalized at an outflow velocity of 6~km~s$^{-1}$. 
The central $\pm 3$~km~s$^{-1}$ part of the spectra 
have been blanked
to avoid contamination from ambient cloud emission.
\label{h2o_co_spec}}
\end{figure}

The velocity information contained in the H$_2$O and CO spectra
offers additional clues on the properties of the 
gas components responsible for the two emissions.
A first comparison between H$_2$O and CO spectra in outflows
was carried out by \citet{fra08},
who used 557~GHz data from SWAS and CO(1--0) data from the FCRAO 14m telescope.
These data represented emission averages over the full extent of the target
outflows 
due to the low resolution of the SWAS observations, and showed that
the H$_2$O lines systematically had more prominent 
wings than the CO(1--0) lines. 
A similar behavior 
has been found by numerous later studies
using different telescopes, spatial resolutions, and (low) $J$ 
values of the CO 
line \citep{bje09,kri12,san12,vas12,nis13}.
\citet{kri12}, in particular, used Herschel 557~GHz data toward 
a sample of 29 YSOs, many of them associated with outflows in our
survey, and compared them with JCMT CO(3--2)
data convolved to the same angular resolution. They found that the
557~GHz outflow line wings were systematically flatter than the CO(3--2)
line wings, and that
557~GHz/CO(3--2) line ratio increased on average by more than one
order of magnitude between the lowest and highest speeds in the
outflow. Similar 557~GHz/CO(3--2)
line ratio increases with velocity 
have been found by \citet{nis13} toward a number
of outflow positions in L1448.

The 557~GHz observations of our survey complement
the YSO-centered observations of \citet{kri12}, since
most of our positions exclude the central object.
For this reason, we
have used our survey data to extend the comparison between 
H$_2$O and CO spectra, and to search for
systematic deviations between the H$_2$O and CO outflow
wing components.
While our H$_2$O data have a lower S/N than the
data from \citet{kri12}, the 557~GHz H$_2$O lines clearly show a
pattern of more prominent outflow wings than the CO(2--1) lines, 
in good agreement with previous studies.
Fig.~\ref{h2o_co_spec} illustrates this pattern with spectra
in logarithmic scale
from the six brightest objects in our sample. 
These spectra have been 
normalized to unity at a velocity of 6~km~s$^{-1}$
away from the ambient speed to ensure that the wing 
comparison is not affected by ambient cloud emission
(which we estimate
extends only $\pm 3$~km~s$^{-1}$ from the systemic velocity).
As can be seen, the H$_2$O wings (in red) are significantly
flatter than the CO(2--1) wings (in blue) at all 
outflow velocities larger than 6~km~s$^{-1}$.
While this pattern is general, the 
difference between the H$_2$O and CO wing slopes
depends on the object, being smallest toward I16293-R and being 
highest
toward HH211-C and CEPE-B. Our survey data, and additional
lower intensity data not shown here, suggest that the difference 
between the H$_2$O and CO outflow slopes may increase as the
the H$_2$O linewidth of the spectrum increases, although higher S/N data
are needed to put this trend on solid ground.

\citet{fra08} interpreted the flatter H$_2$O line wings and the
increase in the H$_2$O/CO  ratio with velocity
as an indication of
an increase in the H$_2$O abundance toward the fastest outflow
gas. This interpretation, however, assumed that both the
H$_2$O and CO emissions arise from the same material,
and that the ratio between the H$_2$O and CO intensities
is proportional to the ratio between column densities.
As discussed before, a number of Herschel observations indicate that the
H$_2$O and low-$J$ CO emissions originate in different
gas components, and  therefore, that the ratio between the H$_2$O
and CO intensities does not correspond to a ratio between 
column densities in the same volume of gas \citep{san12,nis13}.
For this reason, the latter H$_2$O line wings and the
increase in the H$_2$O/CO  ratio with velocity
cannot be interpreted as an abundance effect.
It is more likely that it results from the
H$_2$O-emitting component having a 
significantly larger fraction of fast-moving gas
than the CO(2--1)-emitting gas. 
This difference  could result from the H$_2$O-emitting gas 
representing material that has suffered a
faster shock than the CO-emitting gas, or alternatively,
it could indicate a time evolution
effect, by which the H$_2$O-emitting gas represents 
recently-shocked material that with time will
evolve into the colder and slower CO-emitting component.
More detailed observations involving additional
transitions of both H$_2$O and CO are needed to clarify this issue.

\begin{figure}
\centering
\resizebox{\hsize}{!}{\includegraphics{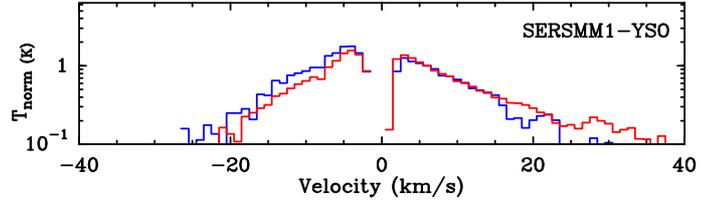}}
\caption{Comparison between H$_2$O 557 (red) and 1670~GHz (blue)
spectra in log scale toward the SERSMM1 YSO illustrating 
their similar outflow wing slopes.
The spectra has been re-centered in velocity and normalized in
intensity as those in Fig.~\ref{h2o_co_spec}.
(Data to be presented in \citealt{mot13}.)
\label{mottram}}
\end{figure}

If the different physical origin of the 557~GHz and CO(2--1) 
emissions 
is associated with a difference in the slope of
their line profiles,
the common origin of the
557 and 1670~GHz emissions suggests that
the two lines should have similar profiles.
As mentioned above, our PACS 1670~GHz 
data do not resolve the emission in velocity, so they
cannot provide information on the shape of the spectral profiles.
Several sub-projects within WISH, however, have carried out
HIFI observations of the 1670~GHz line
toward a number of outflow sources, and these data provide
a limited sample to compare 557 and 1670~GHz line profiles.
\citet{san12} and \citet{vas12}, for example, 
carried out multi-line analysis
of selected positions in the L1448 and L1157 outflows,
and their data show that the 557 and 1670~GHz line profiles are
more similar to each other than to the low-$J$ transitions
of CO, even when observed with telescope beams that differ by a factor
of 9 in area. 
Additional velocity-resolved observations of the 557 and 1670~GHz 
lines will be presented by \citet{mot13}, who observed
five low-mass YSOs, three of them powering outflows included in our
survey. While the \citet{mot13} observations are centered on the
YSO position, and therefore do not sample the same outflow gas
represented
in the 557-1670~GHz intensity correlation of Fig.~\ref{correl}, 
they are the closest data set with which we can check the 
expected similarity between the 557 and 1670~GHz line profiles
in our outflow sample.
As \citet{mot13} show, the 557 and 1670~GHz line profiles 
do in fact look extremely similar. To illustrate it, we
present in Fig.~\ref{mottram} 
a superposition between the 557 and 1670~GHz spectra from
SERSMM1, the brightest source that is common to our sample and
that of \citet{mot13}.
The spectra in the figure have been normalized and presented
as those in Fig.~\ref{h2o_co_spec}, to allow a direct comparison.
As can be seen, the two H$_2$O lines present almost equal wing slopes,
despite having been observed with telescope beams that
differ by a factor 9 in area. This 
good match between the two H$_2$O 
spectra represents a strong confirmation that
the two transitions must originate from the same gas 
component\footnote{The small effect of the
beam size suggests that the emitting region is 
significantly smaller than $39''$.}. 

\subsection{The I(557)/I(1670) intensity ratio}
\label{sec_ratio}

\begin{figure}
\centering
\resizebox{\hsize}{!}{\includegraphics{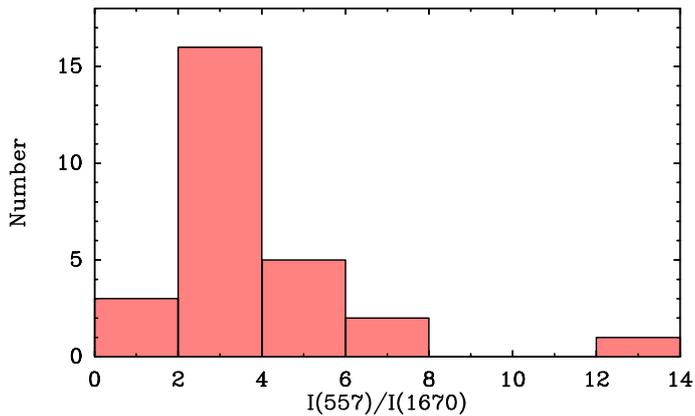}}
\caption{Histogram of the intensity ratio between the 
H$_2$O 557 and 1670~GHz lines. Note the narrow peak between
values 2 and 4 that contains 60\% of the sample.
\label{ratio_obs}}
\end{figure}

An alternative measure of the tight correlation between the
557 and 1670~GHz intensities 
comes from the distribution of the $I(557)/I(1670)$ ratio.
This ratio is more robust than the individual
intensities, since it is less sensitive to
beam dilution and 
the impact of the ambient self absorption, assuming that
the absorption affects the two lines in a similar way (as suggested
by the data from \citealt{mot13}).
For this reason, the $I(557)/I(1670)$ ratio is a better tool
to constraint the physical conditions of the emitting gas, an
issue which will be explored further in Sect.~\ref{sec_nt},
where its value is connected to the gas thermal pressure. 

Fig.~\ref{ratio_obs} shows 
a histogram of the $I(557)/I(1670)$ ratio 
in our outflow sample. 
As can be seen, the histogram presents a narrow distribution,
with 60\% of the objects having ratios between 2 and 4.
This narrow distribution of ratios is a direct consequence
of the tight correlation between the two intensities
seen in Fig.~\ref{correl}, and makes the $I(557)/I(1670)$ ratio
one of the main observables from our outflow survey.

An important property of the $I(557)/I(1670)$ ratio is
that it has a simple interpretation in
terms of line excitation. In the next section, we will
use an LVG radiative transfer analysis to show that
both the 557 and 1670~GHz lines are close to the
optically thin limit. In this limit
the ratio can be written as
$${I(557) \over I(1670)} = {1 \over 3}\; 
e^{53.4 /T_{\mathrm ex}(2_{12}-1_{10})},$$
where  $T_{\mathrm ex}(2_{12}-1_{10})$ is the excitation
temperature in kelvins between the upper levels
of the 1670 and 557~GHz transitions
(the cosmic background radiation has been ignored
due to the high frequencies of the lines and the
high temperature of the gas).

While $T_{\mathrm ex}(2_{12}-1_{10})$ is not a good approximation of
the gas kinetic temperature due to the strong 
subthermal excitation of the H$_2$O molecule, the LVG analysis 
shows that $T_{\mathrm ex}(2_{12}-1_{10})$ is a good
approximation to the excitation temperature of the 
557 and 1670~GHz transitions for the typical
conditions in our sample.
Using the previous formula, we estimate a mean (and median)
$T_{\mathrm ex}(2_{12}-1_{10})$ value of 24~K, and an rms
of 5~K. The small dispersion of $T_{\mathrm ex}(2_{12}-1_{10})$ 
distribution shows that despite our water lines covering two orders
of magnitude in intensity, the emission originates from a
relatively narrow range of physical conditions. 
Determining such conditions in terms of density and temperature
is the goal of the next section.

\begin{figure*}
\centering
\resizebox{15cm}{!}{\includegraphics{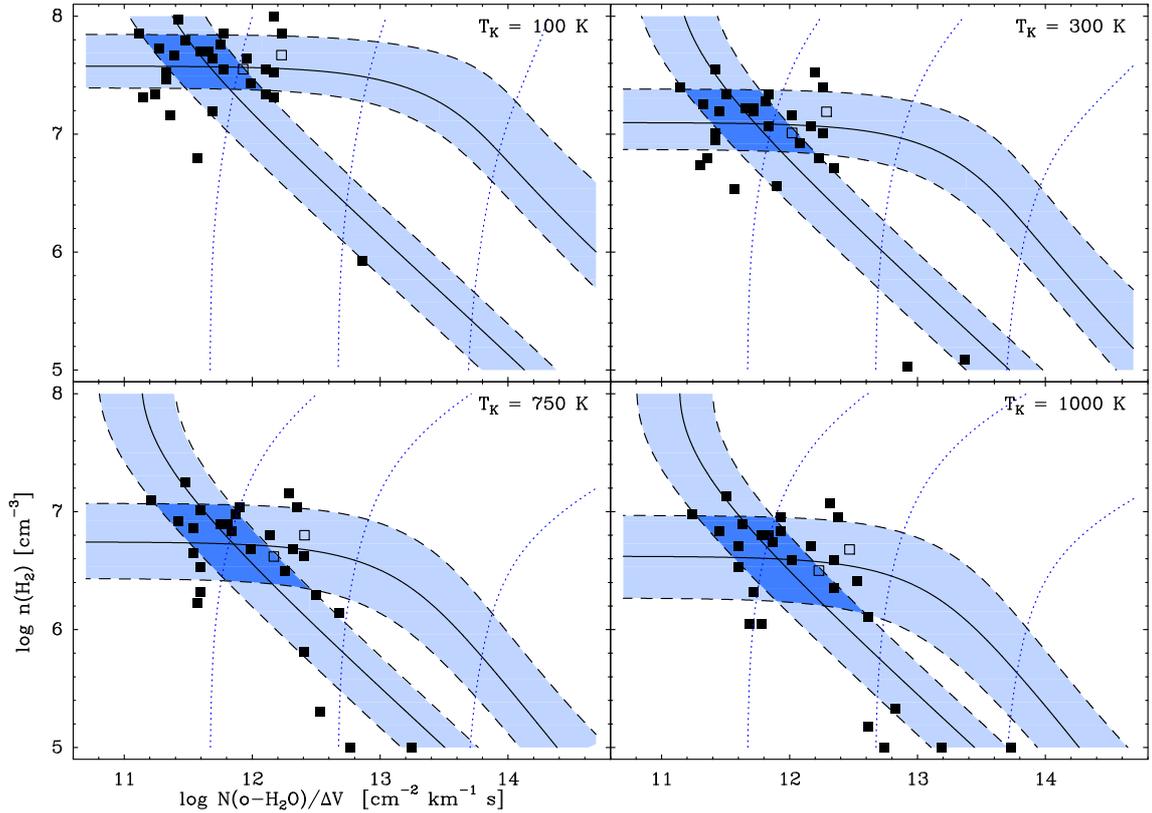}}
\caption{LVG results from the combined analysis of the 557 and
1670~GHz  H$_2$O lines. Each panel corresponds to the choice of
gas kinetic temperature indicated in
the upper right corner. The filled squares represent the best fit
values of
$n({\mathrm{H_2}})$ and $N(\mathrm{H_2O})/\Delta V$
for the outflows in our sample (one point per source), and the open
squares represent fits for L1157-B1 (rightmost point) and
L1448-R4 derived using literature values. 
In all cases, the $n({\mathrm{H_2}})$ and 
$N(\mathrm{H_2O})/\Delta V$ values
have been determined by finding, among a grid of 
more than $10^4$ LVG models, the one that 
best fits the observed $I(557)/I(1670)$ 
ratio (within $39''$) and the peak $I(1670)$
intensity (within $13''$). As discussed in the text, this method
extrapolates the $I(557)/I(1670)$ ratio to a resolution of $13''$ 
based on its approximately constant value over the sample.
The partially horizontal blue-shaded band
is the locus of $I(557)/I(1670)$ values
typical of the sample shown in Fig.~\ref{ratio_obs} (the solid line 
corresponds to a ratio of 3 and the dashed lines to ratios 2 and 4).
The blue-shaded band that runs approximately in diagonal is the locus
between the first and third quartiles of the
observed 1670~GHz peak intensities (solid line is 0.26~K and dashed lines
are 0.12 and 0.47~K).
The blue dotted lines mark the curves of $\tau(1670) = 0.1, 1,$ and
10 ordered by increasing $N(\mathrm{H_2O})/\Delta V$.
The values at the lowest end of the $n({\mathrm{H_2}})$ 
range are upper limits.
\label{lvg}}
\end{figure*}

\section{Physical conditions of the H$_2$O-emitting gas}

\subsection{LVG analysis of the H$_2$O emission}
\label{sec_lvg}

To determine the physical conditions of the gas responsible
for the H$_2$O emission, we
need to solve the coupled equations
of radiative transfer and statistical equilibrium
for the H$_2$O molecule.
To do that, our observations 
only provide two constraints
(the intensities of the 557 and 1670~GHz lines),
and this limits our
search for solutions to those 
having homogeneous gas conditions.
In reality, the
H$_2$O-emitting gas will likely have
internal gradients of density and temperature, so 
our modeling should
be considered as providing average values of the real
gas conditions.

The large range of velocities present in the
HIFI spectra indicates that the emitting gas  
contains strong velocity gradients. 
These gradients decouple the radiation from different 
positions of the cloud, and justify
using the so-called large velocity gradient (LVG) limit
to solve the radiative transfer.
In this limit, the radiative excitation term
of the statistical equilibrium equations
can be treated locally, and this simplifies
enormously the solution
\citep{sob60,cas70,sco74,gol74}.

Our LVG code is based on that presented by \citet{bie93}, and 
assumes that the emitting gas is spherical, since spherical geometry
provides a solution that is intermediate 
among the possible choices of geometry and line broadening
mechanism \citep{whi77}. To include the H$_2$O molecule in
the code, we have added 
the molecular parameters provided by the
Leiden Atomic and Molecular Database (LAMDA,
\citealt{sch05}), which include
the most recent collision rates between H$_2$O and H$_2$ 
\citep{dub06a,dub06b,dub09,dan11,dan10,val08}.

Even using an LVG approximation and assuming homogeneous gas
conditions, three parameters are required to specify the 
solution: 
the gas kinetic temperature $T_{\mathrm k}$, the volume 
density $n({\mathrm{H_2}})$, and the
ratio of the H$_2$O column 
density over the linewidth,  $N(\mathrm{H_2O})/\Delta V$. 
This number of free parameters is larger than the number of our
constraints, so the 
radiative transfer solution is not completely constrained
by the data. Thus, to explore
the full set of possible solutions, we have 
run a series of LVG models fixing each time the gas kinetic
temperature and varying both $n({\mathrm{H_2}})$
and $N(\mathrm{H_2O})/\Delta V$ with logarithmic size steps.
Each of these constant-temperature grids provides a
well-constrained problem in which
the intensities of the two 
H$_2$O lines can be inverted to derive
best-fit values of $n({\mathrm{H_2}})$ and $N(\mathrm{H_2O})/\Delta V$.
This procedure is
illustrated in Fig.~\ref{lvg}, where we present
the results for 4 different grids of kinetic temperature
that range from 100 to 1000~K. 
The 100~K lower limit has been set because
colder models predict densities $\ga 10^8$~cm$^{-3}$,
which seem too high for typical outflow gas.
Temperatures higher than
our 1000~K upper limit are possible, although 
they exceed typical single-temperature
estimates based on H$_2$ emission data (e.g., \citealt{mar09}),
and therefore seem unlikely.
In addition to the outflows of our survey (solid squares),
we present in Fig.~\ref{lvg} the results for 
two well-known outflow positions whose H$_2$O emission has
been studied previously, L1157-B1 and L1448-R4 (open
squares). To obtain these solutions, we have used the intensities
and line ratios
provided by \citet{lef10}, \citet{nis10a}, \citet{san12},
and \citet{nis13}, and we have performed the
same radiative transfer analysis applied to the 
outflows in our sample.
The overlap between all solutions 
shows that our survey outflows 
are not qualitatively different from those
of the two prototypical L1157 and L1448 systems.

As can be seen in Fig.~\ref{lvg}, most best-fit points in each
constant-temperature grid cluster inside a narrow range of
$n({\mathrm{H_2}})$ values, especially for the lowest 
choices of $T_{\mathrm k}$. This clustering of solutions
is a direct consequence of the
narrow range of $I(557)/I(1670)$ ratios found in the
previous section.
To better appreciate this effect, we have plotted in each panel
several lines of constant  $I(557)/I(1670)$ ratio, using as before
values convolved to a $39''$ resolution.
These lines run almost horizontally  for low values of 
$N(\mathrm{H_2O})/\Delta V$ because in this optically
thin regime the excitation is controlled by collisions and therefore
is fixed for each $n({\mathrm{H_2}})$. 
In the optically thick regime (large $N(\mathrm{H_2O})/\Delta V$), 
the lines of constant ratio
bend down toward lower $n({\mathrm{H_2}})$ values because
the contribution from photon trapping lowers the 
density required to achieve a given excitation. 

As can be seen in the figure, most outflow points 
lie inside the horizontal
blue band bounded by line ratios 2 and 4
(dashed lines), which has a 
width of 0.5-0.6 dex in density. This means that
most solutions deviate less than a factor of 2 from
the density corresponding to 
the median ratio of 3 (solid line).
Although the horizontal blue band contains most of
the LVG points, a number of solutions lie at significant
lower densities.
These points correspond
to $I(557)/I(1670)$ ratios
that exceed 4, and their
large spread in the plot
reflects the strong sensitivity of the derived gas
density to the value of the line ratio when it is larger 
than 4.

\begin{figure*}
\centering
\resizebox{\hsize}{!}{\includegraphics[height=14.5cm]{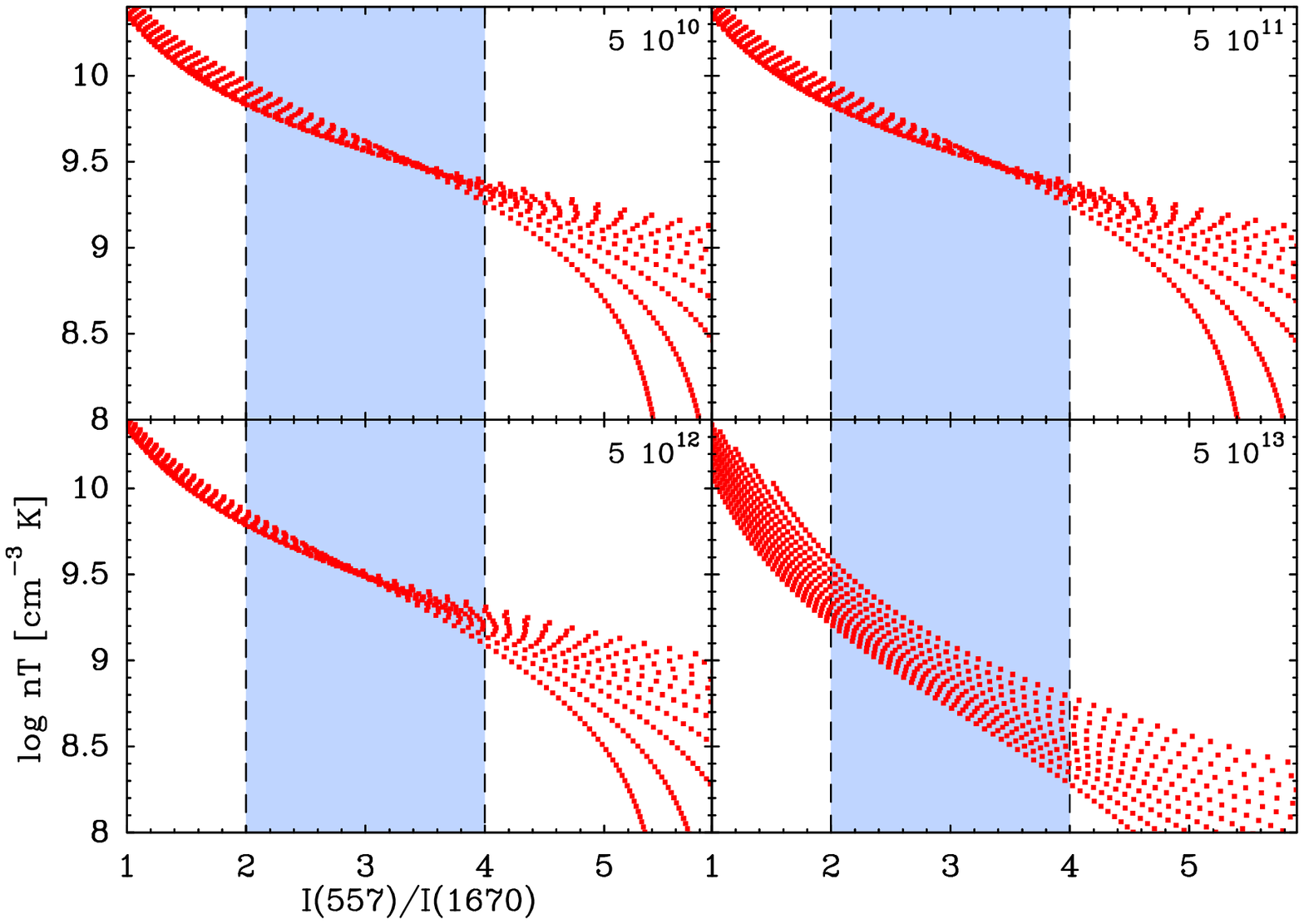}
\includegraphics[height=14.5cm]{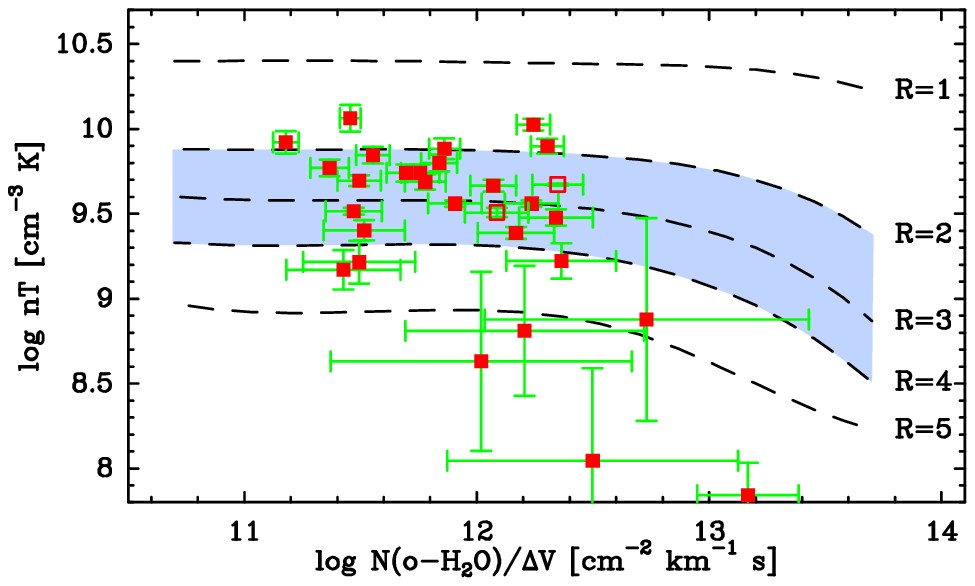}}
\caption{{\bf Left:} gas thermal pressure $nT$ vs. $I(557)/I(1670)$ 
as determined by 
a series of LVG models. Each panel summarizes the result from
more than 1000 LVG models of different density and temperature
(see text), together with a
constant value of $N(\mathrm{H_2O})/\Delta V$ indicated
in the upper right corner (units are cm$^{-2}$km$^{-1}$s).
The blue-shaded region marks the interval of ratios between 2 and 4
that contains 60\% of the outflow sample.
{\bf Right:} Thermal pressure $nT$ vs. $N(\mathrm{H_2O})/\Delta V$
for the outflows in our sample. Each point
represents an outflow (open squares correspond to L1157-B1 and L1148-R4), 
and the mean values and error bars
have been determined from the LVG results shown in
Fig.~\ref{lvg}. The dashed lines labeled with
$R$ ($=I(557)/I(1670)$) have been derived from the models
in the left panels. Again, the region
between $R$ values of 2 and 4 is shaded in blue.
\label{nt}}
\end{figure*}

While the value of the $I(557)/I(1670)$ ratio controls the 
best-fit volume density
in the LVG model, the absolute line intensities control
the derived value of $N(\mathrm{H_2O})/\Delta V$.
This is illustrated in Fig.~\ref{lvg} 
by the lines of
constant 1670~GHz intensity.
These lines 
cross the LVG grid diagonally from top to bottom,
and tend to run almost vertically at high densities,
since the levels are thermalized.
In the optically thin regime (low $N(\mathrm{H_2O})/\Delta V$)
the constant-intensity lines intersect the curves of
constant $I(557)/I(1670)$ ratio at a single point, and this means that 
for a given line ratio, different I(1670) intensities
correspond to solutions of fixed $n({\mathrm{H_2}})$ but
different $N(\mathrm{H_2O})/\Delta V$.

If the column density estimate depends sensitively
on the observed line intensity, our results
are potentially sensitive to beam-dilution effects.
In Sect.~\ref{sec_size} we saw that the PACS 
maps indicate typical emission sizes of $20''$,
which is significantly smaller than the HIFI
$39''$ resolution used estimate the $I(557)/I(1670)$ ratio.
For this reason, an LVG analysis 
using $39''$-beam intensities
will necessarily underestimate the
value of $N(\mathrm{H_2O})$.
To mitigate this problem, we have
carried out our LVG analysis with the
unconvolved 1670~GHz intensities, which have
an intrinsic resolution of $13''$
and are unlikely to be strongly beam diluted
(Sect.~\ref{sec_size}). In addition, for
each source we have chosen the 
peak value of the 1670~GHz line,
which 
maximizes the H$_2$O column density estimate.
Of course, self-consistency 
requires that we also use 
557 GHz line intensities with
$13''$ resolution,
instead of the $39''$ HIFI beam.
As no high-resolution data exist, we have assumed
that the $I(557)/I(1670)$ ratio in the $13''$ PACS beam is the same
as in the $39''$ beam.
This assumption is supported by the almost constant
value of the line ratio
in the sample, which suggests that the ratio is 
independent
on the source distance and on how well centered on the
emission peak our 557~GHz observations were,
and therefore, that
it varies little inside the mapped region.
Thus, our LVG analysis can be thought as constrained by
two independent measurements:
the $I(557)/I(1670)$ ratio determined with a $39''$ beam and extrapolated
to $13''$, and the intensity of the 1670 GHz line
truly measured with $13''$ resolution. 

As Fig.~\ref{lvg} shows, typical $N(\mathrm{H_2O})/\Delta V$ values in
our sample are
around 10$^{12}$~cm$^{-2}$~km$^{-1}$~s, with few
points exceeding 10$^{13}$~cm$^{-2}$~km$^{-1}$~s.
These values represent the peak column density for
each source, since they were estimated using the
peak 1670~GHz intensity.
Other positions of each source lie horizontally to
the left in the LVG diagrams because they have
the same line ratio (assumed constant in each source)
and a lower 1670~GHz intensity.
The relatively low $N(\mathrm{H_2O})/\Delta V$ values
we derive
reinforce the idea that the H$_2$O emission 
cannot be optically thick.
Indeed, the lines of constant $\tau(1670)$ in Fig.~\ref{lvg}
(blue dotted lines)
indicate that most points have values below 1,
and that moving those points into the optically
thick regime would require
multiplying most peak intensities
by factors of 5-10. This
factor seems larger than expected
from dilution effects given the source sizes
estimated in 
Sect.~\ref{sec_size}

\subsection{The 557/1670 ratio and the gas thermal pressure}
\label{sec_nt}

The LVG analysis
illustrates how our observations cannot
constrain completely the physical conditions
of the H$_2$O-emitting gas.
The data can be fitted with a solution 
where the gas has a relatively low temperature (100~K)
together with a high density ($\approx 4\times 10^7$~cm$^{-3}$),
or has 
a higher temperature (1000~K) and a lower 
density ($\approx 4\times 10^6$~cm$^{-3}$).
Both extreme solutions, and many other
in between, produce the same level of
excitation consistent with the
observed $I(557)/I(1670)$
ratio. 

The opposite role that density and temperature play 
in the LVG solution suggests that their product
can be determined better than each individual quantity.
This $n(\mathrm{H}_2) T_k$  product ($nT$ hereafter)
corresponds to the thermal pressure of the gas ($P/k$), 
and is a useful parameter to
constrain shock models.
To explore how well it can be determined from
our data,
we have run series of more than 1000
LVG models fixing each time the column density and 
covering with a fine grid both the density 
($10^5$ to $10^8$~cm$^{-3}$ with logarithmic step of 0.03)
and the temperature (100 to 1000~K 
with a logarithmic step of 0.1).
For each model, the derived intensity
of the 557 and 1670~GHz lines
has been used to estimate the $I(557)/I(1670)$ ratio, 
and scatter plots of $nT$ vs $I(557)/I(1670)$
are presented in the left panel of Fig.\ref{nt} for
four different column density values.
As can be seen, 
there is a tight correlation
between $nT$ and 
$I(557)/I(1670)$ when the line ratio is lower than 5 and 
$N(\mathrm{H_2O})/\Delta V < 5\times 10^{13}$~cm$^{-2}$~km$^{-1}$~s,
which are conditions typical of the outflow
data. This means that the observed line ratio 
can be used to constrain the gas pressure,
even if we cannot distinguish
between the high and low temperature solutions
in the LVG analysis.

To determine the gas pressure in
each object of our sample, 
we have used the four LVG solutions shown in Fig.~\ref{lvg}
and estimated the
mean and dispersion 
values of $nT$ and $N(\mathrm{H_2O})/\Delta V$.
The results are summarized in Table~\ref{summary} and 
plotted in the right panel of Fig.~\ref{nt}.
As expected, the dispersion  in
$nT$ is relatively small for line ratios $R < 5 $ ($< 0.2$~dex),
which include 75\% of our sample.
Typical $n T$ values exceed $10^9$~cm$^{-3}$~K,
and the pressure corresponding to points with $R=3$
(the median line ratio of our sample) 
and typical H$_2$O column densities
is $4\times 10^9$~cm$^{-3}$~K.

The high gas pressures derived with our analysis 
are consistent with the idea that the 
H$_2$O-emitting gas has been
compressed by a strong shock. To determine the 
nature of such a shock, we first estimate the
pressure increase with respect to pre-shock
conditions.
Since most of our outflow positions
lie at some distance from the driving
YSO, we assume pre-shock densities
and temperatures typical
of the cloud gas that surrounds the dense cores,
which means 
$T= 10$~K and $n = 10^4-10^5$~cm$^{-3}$
(e.g., \citealt{ber07}).
These values imply that the pre-shock 
pressures were
in the range $10^5-10^6$~cm$^{-3}$~K,
and therefore, that the pressure enhancement 
by the shock has been
on the order of $10^4$. 

A pressure enhancement of $10^4$ seems uncomfortably
high for a number of shock models,
especially those of C type. In these shocks,
the gas compression is significantly limited by the
contribution from the magnetic field,
and this leads to a relatively small
gas pressure jump.
This can be seen in Fig.~1 of
\citet{flo10}, that
provides detailed 
density and temperature profiles 
for a number of C-shock models.
By simply multiplying the density
and temperature values in these profiles, we
estimate that the gas
pressure jump in C-type shocks is not
expected to exceed a value of 
500 even for shock velocities of 40~km~s$^{-1}$
(the highest considered by the authors).
Pressure jumps for shock velocities of
20~km~s$^{-1}$, which are closer to 
the total velocity extent we find in
the H$_2$O lines, are typically
on the order of 200, which is much smaller
than the $10^4$ factor we derive from the observations.
C-type shock models seem therefore inconsistent with 
the observed pressure jumps determined
from the H$_2$O data.

J-type shocks,
where a weak magnetic field plays no 
dominant role in the kinematics, 
provide a better 
alternative to explain the observed pressure jumps.
Simple analysis of these shocks using the
Rankine-Hugoniot jump conditions shows that
for high velocities, the
post-shock over pre-shock pressure ratio is
proportional to the Mach number squared 
(e.g., \citealt{shu92}).
A shock velocity of 20~km~s$^{-1}$
corresponds approximately to a
Mach number of 100 for 10~K gas, so
the expected initial to final
pressure ratio in this type of J-shock is about
$10^4$, similar to what we derive from the observations.
The more detailed shock models presented by 
\citet{flo10} (bottom panels of their Fig.~1), 
confirm the simple analytic theory, and show
pressure jumps close to $10^4$ for a shock speed
of 20~km~s$^{-1}$ and pre-shock H$_2$ 
densities of either $10^4$ or $10^5$~cm$^{-3}$.
The high pressure jumps derived by our 
analysis, therefore, strongly favor J-type shocks over 
C-type shocks as the type of disturbance responsible for
the physical conditions of the H$_2$O-emitting gas in outflows.
If the H$_2$O component coexists with the H$_2$ gas responsible for
emission seen in the Spitzer IRAC images
(as suggested in Sect.~\ref{sec_h2o_h2}), our results imply that 
J-type shocks must also be responsible for the 
H$_2$ emission seen at near and mid IR wavelengths
(see also \citealt{nis10b}).

\section{H$_2$O abundance in the outflow gas}

\subsection{H$_2$O abundance in the warm outflow component}
\label{sec_abu}

The abundance of H$_2$O in the shocked outflow gas
is a critical parameter to test chemical models.
Over the years, a number of authors
have estimated  the abundance of 
H$_2$O in low-mass outflows
using data from different telescopes, such as ISO, SWAS, Odin, and Herschel 
(e.g., \citealt{lis96,nis99,gia01,fra08,bje09,lef10,kri11,vas12,san12,nis13}).
Unfortunately, the derived values cover
a very wide range, 
from about $10^{-7}$ to $10^{-5}$, and no consensus exists
on what the ``typical'' H$_2$O abundance in an outflow is.

It is possible that the large range of H$_2$O abundances 
arises from true chemical differences between the outflows, or
from differences in the abundance of the various
temperature components in the shocked gas.
Still, a significant part of the dispersion
seems to result from differences in the analysis
used to derive the abundance. Broadly speaking, two main issues 
have contributed to the multiplicity of H$_2$O abundance 
estimates.
On the one hand, some estimates have used low-$J$
transitions of CO to infer the H$_2$ 
outflow column density from which the H$_2$O abundance
is determined, 
while other estimates have used
direct determinations of the H$_2$ column density 
from emission at near or mid IR wavelengths. 
As discussed in Sects.~\ref{sec_hh211_cepe} and
\ref{sec_h2o_co}, the low-$J$ transitions of CO trace a
cold outflow component than does not coexist
with the H$_2$O-emitting gas, and as a result, H$_2$O abundance 
determinations based on low-$J$ CO data are likely to be in error.
The mid-IR H$_2$ emission, on the other hand, seems closely-connected with 
the H$_2$O-emitting gas (Sect.~\ref{sec_hh211_cepe}), and therefore 
represents a more reliable tracer of the outflow column density
responsible for H$_2$O. In this section, we will therefore
use this mid-IR H$_2$ emission as the reference for the
H$_2$O-abundance determination 
(see \citealt{vas12} and \citealt{san12} for a similar approach).

The other cause for the dispersion of
H$_2$O abundance values in the literature 
is the diversity of radiative transfer solutions
proposed from the H$_2$O emission. Large H$_2$O abundance values tend to
be associated with optically thick solutions that infer large
H$_2$O column densities, while low abundance estimates 
result from solutions where the optical depth of the
H$_2$O emission is low or moderate
(see \citealt{kri11} for a comparison between the two
different approaches in the case of the L1448 outflow).
As discussed above (\ref{sec_lvg}), our LVG 
analysis suggests that the H$_2$O emission
from the outflows 
in our sample is optically thin or has at most moderate optical depth. 
This means that
our H$_2$O abundance estimate is expected to favor 
values near the low end of the published range.

If the H$_2$O emission has at most moderate
optical depth, 
the intensity of the 1670~GHz line from any 
object must be proportional
to the column density of its emitting H$_2$O.
This means that 
the linear relation between $I(1670)$ and IRAC4 intensities found in
Sect.~\ref{sec_h2o_h2} (and illustrated in Fig.~\ref{irac4_pacs})
must translate into a similar relation between 
H$_2$O and H$_2$ column densities. From this relation,
it should be possible to derive an H$_2$O abundance value
that is representative of our outflow sample.
Of course, the $I(1670)$ vs IRAC4 correlation has
significant scatter, and our 
radiative transfer analysis has a number of 
uncertainties. This allows for some 
scatter in the
H$_2$O abundance of the different outflows.
Still, the scale of this scatter should be on the 
order of a factor of a few, and not 
the two orders of magnitude seen in the literature.

Before proceeding with the analysis, it is important to 
check the consistency between the 
treatments of the H$_2$O and H$_2$  emission.
H$_2$ radiative transfer solutions often result in low
density estimates, as illustrated by the less than 
$10^4$~cm$^{-3}$ values derived by \citet{neu09} from
Spitzer observations. This is of course much lower than our
$> 10^6$~cm$^{-3}$ estimate from the
H$_2$O observations, and brings into question whether the
H$_2$O and H$_2$ emissions can be reproduced with
the same physical conditions (a necessary requirement
if the emissions coexist).
That this is the case has been recently shown by the
\citet{gia11}, who have re-analyzed
the same H$_2$ rotation lines studied by  \citet{neu09}, but 
this time complementing them with  
vibrational transitions. This new analysis  has 
increased the original density estimate of the H$_2$-emitting gas
to values consistent with those derived from our H$_2$O analysis,
showing that it is possible to interpret both emissions
with a consistent set of gas conditions.

If the H$_2$O and H$_2$ emissions arise from
the same volume of gas, 
to calculate an outflow-wide estimate of the 
H$_2$O abundance we need to 
calculate  the proportionality factors
between the $I(1670)$ and IRAC4 intensities and the corresponding
H$_2$O and H$_2$ column densities.
As a first step, 
we derive the conversion factor between
the IRAC4 intensity and the 
H$_2$ column density. We do so following
the methodology of \citet{neu08}, who
have shown that IRAC 
observations of shocked gas can be reproduced 
assuming that the emitting material has a distribution 
of column densities that depends on
temperature as a power law with the form $T^{-\beta}$.
The $\beta$ parameter is typically
4 for bipolar outflows,
and the power law distribution seems valid
approximately between 300 and 4000~K
\citep{nis10b,gia11}.
Our analysis of the IRAC4 emissions makes
therefore use of these literature values,
together with a gas density
of $5\times 10^6$~cm$^{-3}$ \citep{gia11}
and an ortho-to-para ratio of 2.2 for H$_2$
\citep{nis10b}, and predicts 
the emission of the different H$_2$ rotation lines
for temperatures between 300 and 4000~K.
For this we have used the 
LVG code, this time with 
the H$_2$-H$_2$ collision rates
from \citet{flo99} (as provided by the BASECOL database,
see
\citealt{dub06b}) and the Einstein A coefficients for H$_2$ 
from \citet{wol98}.
By integrating the contribution from all temperatures,
and assuming that the $S$(4) and $S$(5) transitions
contribute to the IRAC4 intensity with the weights
determined by \citet{neu08}, we have calculated a relation between
H$_2$ column density and IRAC4 intensity with the form
$$ N(\mathrm{H}_2)\; [\mathrm{cm}^{-2}] = 4.5\times 10^{19} \;  I^*(IRAC4) \;
[\mathrm{MJy~sr}^{-1}] $$
where, as in Sect.~\ref{sec_h2o_h2}, $I^*$(IRAC4) is
the extinction corrected IRAC4 intensity.
To test this relation, we have applied it to the
red lobe of L1448 and the blue lobe of BHR71, for which 
more accurate determinations have been presented by
\citet{gia11}. Reading the color scale in 
Figs.~4 and 5 from these authors, we estimate that
our analytic N(H$_2$) estimates agree with the more accurate values
within 20\%.

While the above column-density-intensity
relation is valid for a mix of gas with temperatures
between 300 to 4000~K, a 
single-temperature analysis of the H$_2$
emission shows that it is equivalent to
the relation for isothermal gas at about 600~K.
This suggests, that the IRAC4 emission is dominated by
the low end of the temperature distribution, which 
should not be surprising
given the steep temperature dependence
of the column density implied by the
$\beta=4$ exponent.

To calculate now the conversion factor between the 1670~GHz
line intensity 
and the H$_2$O column density, we need an analysis that is consistent
with that of H$_2$. This means that we cannot
apply the single-temperature treatment used in Sect.~\ref{sec_lvg},
but that we have to assume  a 
distribution of H$_2$O column densities that also follows 
a $T^{-\beta}$ law with $\beta=4$. Also, we have to
assume the same gas volume density
of $5\times 10^6$~cm$^{-3}$ that
was used for H$_2$.
With these values, plus an H$_2$O ortho-to-para ratio
of 3 (e.g., \citealt{her12}), 
we have run 
a series of LVG models covering the temperature
range from 300 to 4000~K, and we have integrated the 
resulting intensity of the 1670 GHz line 
weighting it by the temperature-dependent 
column density distribution. 
The resulting relation has the form
$$ N(\mathrm{H_2O})\; [\mathrm{cm}^{-2}] = 4.8\times 10^{12} \;  
I(\mathrm{1670~GHz}) \;
[\mathrm{K~km~s}^{-1}], $$
where the column density refers to the total (ortho + para)
value. 

If we again compare the multi-temperature relation with 
a single-temperature analysis, we find 
that it is equivalent to that of an isothermal
gas at about 450~K. This temperature is similar 
to the 600~K derived from the H$_2$ analysis,
confirming the idea that both emissions are 
dominated by gas at the low
end of the temperature distribution.
This should not by surprising given the steep power-law assumed
for the distribution of H$_2$ column density with temperature,
which implies that, for example,
less than 7\% of the gas is at temperatures higher than 750~K.
It indicates that the H$_2$O abundance
estimate we are about to derive
is dominated by gas at around 500~K.

The single-temperature H$_2$O estimate also allows
a consistency check with the pressure analysis of
the previous section. Combining the derived 450~K
with the assumed density of $5\times 10^6$~cm$^{-3}$,
we derive a gas pressure
$\log (nT) = 9.35$, which according to Fig.~\ref{nt}
is again inside the range of observations and corresponds 
to $I(557)/I(1670) = 4$. These numbers show that
the multi column density analysis of H$_2$O is 
consistent both with the analysis of the H$_2$ emission 
and with the analysis of the 557 and 1670~GHz
intensities presented
in the previous sections.

Combining the above intensity-column density relations for H$_2$ and
H$_2$O with the IRAC4-$I(1670)$ correlation found in Sect.~\ref{sec_h2o_h2},
we derive an approximate H$_2$O abundance of $3\times 10^{-7}$
for the gas responsible of the observed 557 and 1670~GHz emission.
This value has an uncertainty level
of at least a factor of 2, which corresponds to the
0.3 dex rms level in the H$_2$O column densities 
for objects with $R=4$ (Fig.~\ref{nt}). Other sources
of uncertainty related to the radiative transfer
of H$_2$O and H$_2$ are possible and can add to
the error budget, but they cannot be easily
quantified without the observation of additional
H$_2$O lines. In any case, our estimate clearly favors
a relatively low abundance value for H$_2$O, 
as expected from the optically thin analysis
(and in line with
recent estimates for individual outflow sources, like
L1157 by \citealt{vas12} and VLA1623 by \citealt{bje12}).

Our low H$_2$O abundance is in clear conflict
with the expectation from C-type shock models, which are often
used in the analysis of the H$_2$O emission from outflows. These
models consistently predict H$_2$O abundances in excess of $10^{-5}$ 
due to an almost complete conversion of oxygen into H$_2$O 
\citep{kau96,flo10}. This value exceeds our
derived abundance value by more than one order of magnitude
and therefore is excluded by our analysis.
As mentioned before, however, C-type shocks seem
already inconsistent with the high gas pressure 
inferred for the H$_2$O emitting gas, so their
failure to match the observed H$_2$O abundance
should not be considered surprising.
The alternative J-type shock models have
unfortunately received much less attention.
\citet{flo10} showed that these models
do in fact predict lower  H$_2$O abundances
due to its destruction by collisions with atomic hydrogen,
although their Fig.~3 suggests that 
a large enough abundance decrease only occurs in 
post-shock gas that is too cold to be consistent
with our observations. Additional contribution
from UV radiation in fast shocks may help decrease the 
post-shock abundance of H$_2$O, according to \citet{neu89}.
Further work on destruction mechanisms of H$_2$O
in shocks is clearly needed 
to understand the low abundances derived from the data.

\subsection{H$_2$O abundance in the cold outflow component}
\label{sec_cold}

Throughout the paper, we have distinguished between two outflow
components, a relatively cold one traced by the low-$J$ CO
transitions and a warmer one traced in H$_2$.
The H$_2$O emission observed by Herschel arises from the
warm outflow component, so the
H$_2$O abundance estimated in the previous
section only applies to this
higher excitation part of the flow.
In this section we investigate how much 
H$_2$O can be hidden in the cold component
of the outflow, and how the H$_2$O abundance in
this component can be further investigated with observations.

Since no model of
the different outflow components exist in the literature,
our analysis will use the simplest 
assumptions consistent with the data.
We simplify the outflow gas structure as
consisting of two components, one
responsible for the low-$J$ CO emission and other responsible for the
H$_2$O emission. For the low-$J$ CO-emitting component, we assume
a gas temperature of 30~K and a volume density of $10^5$~cm$^{-3}$,
which are values typically derived from low-$J$ molecular
transitions (e.g., \citealt{taf10}). For the 
H$_2$O-emitting gas, we use the values derived in the previous 
section, i.e., a representative temperature of 
450~K and a density of $5\times 10^6$~cm$^{-3}$.

As a preliminary check, we make sure that our
model is consistent with the idea that the low-$J$ CO
emission is dominated by the cold component and that
the contribution from the warm gas is negligible.
To test this, we first calculate the
CO column density of the warm outflow.
From our analysis of the HIFI data, we derive a 
mean linewidth is 16~km~s$^{-1}$,
and assuming that the linewidths of the 557 and
1670~GHz are equal, we estimate a
the median 1670~GHz line brightness of 0.3~K.
These values, 
together with the relation from the previous section, imply
a typical column density of of warm H$_2$O of $2.3\times 10^{13}$~cm$^{-2}$.
If we now assume an H$_2$O abundance
of $3\times 10^{-7}$ (Sect.~\ref{sec_abu}) together with a standard 
CO abundance of $8.5\times 10^{-5}$ \citep{fre82}, we derive that
the CO column density in the warm component is $7\times 10^{15}$~cm$^{-2}$.
For the assumed temperature of 450~K and 
density of $5\times 10^6$~cm$^{-3}$, 
an LVG model predicts that the
CO(1--0) and CO(2--1) intensities must be $<0.1$~K.
Such intensities are clearly smaller than the $\sim 3$~K
observed with the IRAM 30m telescope,
in agreement with the expectation 
that little low-$J$ CO emission comes from the warm outflow gas.

We now investigate the H$_2$O content of 
the cold outflow component. First, we derive the
H$_2$ column density of this part of the outflow using our 
complementary IRAM 30m data, which show 
typical CO(2--1) intensities of 
3~K and typical linewidths of 10~km~s$^{-1}$.
Making use again of the LVG code, this time for 
a temperature of 30~K and a volume density of 
$10^5$~cm$^{-3}$ (together with the previous CO abundance),
we estimate that the cold outflow gas has a typical
H$_2$ column density of $2.1\times 10^{20}$~cm$^{-2}$ 
in a $13''$ beam.

An observational constraint to the 
H$_2$O abundance in the cold outflow is that it
should remain undetected in our Herschel observations.
We thus explore how much H$_2$O can remain hidden
in the cold gas. As a first guess,
we assume an abundance level equal to that found
in the warm
component ($3\times 10^{-7}$). With this value and
the LVG model, we
predict intensities 
of 0.85~K~km~s$^{-1}$ for the 557~GHz line
and 0.01~K~km~s$^{-1}$ 
for the 1670~GHz line.
These values can be compared with the results from
our Herschel observations summarized in 
Table~\ref{summary}. As can be seen, the 
predicted 557~GHz intensity from the
cold outflow component is almost 6 times smaller than
the mean observed value, while the predicted
1670~GHz intensity is two orders of magnitude
lower than observed.
These lower-than-observed values indicate that
the emission from the cold outflow will be
overwhelmed by the emission from
the warm component, and therefore 
likely missed in an observation. As a result, it seems
possible to hide an H$_2$O abundance of $3\times 10^{-7}$
in the cold outflow gas for a significant number of 
outflows from our sample. 
(Outflows with weak H$_2$O emission, such as L1551,
likely can only hide lower abundances.)

While the above estimate suggests that it is
possible to hide an H$_2$O abundance level of
$3\times 10^{-7}$ in the cold component of some 
outflows, it seems unlikely
that a much higher value can remain undetected.
The previous LVG solution for a typical (cold) outflow
component predicts an optical depth of 1.1 for
both the 557 and 1670~GHz lines. 
This value, together with the
expected low excitation temperature
of the two transitions
($\sim 5$~K), indicates 
that a higher H$_2$O abundance in the cold gas
will cause a noticeable self absorption feature
in the spectrum. Such a self absorption
should be easily distinguishable
from the narrow ambient absorption feature
seen in the spectra,
since it should appear as a
relatively broad, wing-like dip in the
spectra of both 557 and 1670~GHz lines.

Our 557~GHz HIFI data do not show evidence for
broad self-absorptions in the spectra
(Fig.~\ref{summary-fig}),
and this suggests that abundance values much 
larger than $3\times 10^{-7}$ are unlikely for the cold
outflow gas in the objects of our sample.
Some H$_2$O-bright outflows, however, do present
features that could be indicative of 
cold H$_2$O. \citet{vas12} and \citet{san12} have
shown that in L1157-R and L1448-R4,
the spectra from transitions connected with the ground
state of both ortho and para-H$_2$O present a deficit
of emission at low velocities compared with the spectra
from excited levels. Whether these features result from 
self-absorption by cold H$_2$O or from an entirely 
different process
needs to be assessed with detailed multi-transition 
spectral modeling. Such an investigation 
can potentially provide additional constrains on
the H$_2$O abundance in the cold outflow gas,
and thus help complete the analysis of H$_2$O in
outflows presented here.

\section{Summary}

We have carried out a survey of H$_2$O emission toward a sample
of mostly young bipolar outflows using the Herschel Space 
Observatory. This survey was part of the ``Water In
Star-forming regions with Herschel''
(WISH) project, and combined HIFI
observations of the 557~GHz line with PACS footprints
of the 1670~GHz line toward typically two positions
in about 20 outflows. From the analysis of these data, together
with complementary CO(1--0) and CO(2--1) observations carried out with the
IRAM 30m telescope and archive Spitzer/IRAC data, we have reached the 
following main conclusions:

\begin{enumerate}

\item The spatial distribution of the 
H$_2$O emission tends to resemble
the distribution of IRAC-derived H$_2$
emission, while it differs from the distribution
of both CO(1--0) and CO(2--1).
This dichotomy of distributions suggests that
H$_2$O traces a gas component closely connected
with the H$_2$-emitting gas (at hundreds of
kelvins) and distinct from the gas producing
the low-$J$ CO emission (at tens of kelvins)
(Sect.~\ref{sec_hh211_cepe}).

\item In addition to spatial coincidence, the H$_2$O and H$_2$
emissions correlate in intensity. We find an approximately
linear correlation between the intensities of the 1670~GHz
emission traced with PACS and the H$_2$-dominated intensities
observed by the different Spitzer IRAC bands
(Sect.~\ref{sec_h2o_h2}).

\item The analysis of the PACS footprint maps indicates that
the H$_2$O emission is concentrated but not point-like.
It often consists of a combination of bright peaks and
extended emission, and the deconvolved 
typical emission size is around $20''$
(Sect.~\ref{sec_size}).

\item Most HIFI 557~GHz spectra present outflow
wings together with ambient-speed absorption
features. 
The wing shape of the spectra indicates that while 
some outflow H$_2$O emission originates
in high velocity gas, most of the emission comes from
relatively slow material. Such distribution 
contrasts with the expectation
from plane-parallel shock models, which predict spectra
having a narrow emission feature at the highest speeds
(Sect.~\ref{sec_lineshape}).

\item  There is a tight correlation between the integrated 
intensities of the 557 and 1670~GHz lines 
over two orders of magnitude, indicating that the two emissions
arise from the same volume of outflow gas. On the
other hand, any correlation between the 
557~GHz and CO(2--1) integrated intensities is
weak at most. This is consistent with 
the two
emissions arising
from different components of the outflow gas
(Sect.~\ref{sec_int_corr}).

\item In agreement with previous work, we find that 
the H$_2$O 557~GHz lines have flatter outflow wings
than the low-$J$ CO transitions. We interpret this effect
as a consequence 
of the different kinematic properties of the gas 
responsible for the two emissions
(Sect.~\ref{sec_h2o_co}).

\item Combining the analysis of the 557 and 1670~GHz 
lines, we find a relatively narrow range of intensity
ratios, with most objects lying between values 2 and 4. 
The observed line ratios suggest H$_2$O excitation temperatures on the
order of 25~K (Sect.~\ref{sec_ratio}).

\item 
An LVG analysis of the 557 and 1670~GHz
lines shows that our set of two transitions
is not enough to constrain all the physical
conditions of the H$_2$O-emitting gas.
It seems equally possible to fit the data
with solutions that are relatively cold (100~K) and dense
($4\times 10^7$~cm$^{-3}$), solutions that are 
warm (1000~K) and less dense ($4\times 10^6$~cm$^{-3}$),
and a number of intermediate values. In all
cases, the models are consistent with the
emission being optically thin (Sect.~\ref{sec_lvg}).

\item While our data cannot constrain separately 
the density and temperature of the H$_2$O-emitting gas,
they determine with little dispersion 
the product, which is proportional 
to the gas pressure.
The pressure values we derive indicate that the H$_2$O-emitting
outflow component
is over-pressured with respect to the ambient cloud
by factors on the order of $10^4$.
Such high levels of compression seem inconsistent 
with C-type shocks, and suggest that J-type shocks
are responsible for the observed gas conditions 
(Sect.~\ref{sec_nt}).

\item
Combining the observed correlation between 
PACS and IRAC intensities with the excitation conditions derived 
from the LVG analysis, we derive a typical
H$_2$O abundance of $3\times 10^{-7}$ for the gas responsible 
of the observed transitions. 
While uncertain by a factor of a few, this value is
significantly lower than standard abundance predictions from
C-type shocks. J-type shock models may be able to
fit the observations, although more work on H$_2$O
destruction mechanisms in this type of shocks are still 
needed (Sect.~\ref{sec_abu}).

\item
Our derived H$_2$O abundance corresponds to the 
warm ($\sim 500$~K) component of the outflow gas.
A simple model suggests that a similar 
abundance level could remain hidden 
in the cold component of a number of outflows.
Further progress investigating the abundance of
H$_2$O in this cold outflow component could be made
searching for broad self-absorption components
in H$_2$O spectra (Sect.~\ref{sec_cold}).

\end{enumerate}

\begin{acknowledgements}
We thank Joseph Mottram for providing us with the SERSMM1 data of Fig.~\ref{mottram}
prior to publication, and 
Doug Johnstone and Susanne Wampfler for their
useful comments and suggestions.
We also thank an anonymous referee and Malcolm Walmsley for comments that 
helped clarify the presentation.
HIFI has been designed and built by a consortium of institutes and university 
departments from across Europe, Canada and the United States under the leadership 
of SRON Netherlands Institute for Space Research, Groningen, The Netherlands and 
with major contributions from Germany, France and the US. Consortium members 
are: Canada: CSA, U.Waterloo; France: CESR, LAB, LERMA, IRAM; Germany: KOSMA, 
MPIfR, MPS; Ireland, NUI Maynooth; Italy: ASI, IFSI-INAF, Osservatorio Astrofisico 
di Arcetri-INAF; Netherlands: SRON, TUD; Poland: CAMK, CBK; Spain: Observatorio 
Astron\'omico Nacional (IGN), Centro de Astrobiolog\'{\i}a (CSIC-INTA); 
Sweden: Chalmers University of Technology - MC2, RSS \& GARD; Onsala Space 
Observatory; Swedish National Space Board, Stockholm University - Stockholm 
Observatory; Switzerland: ETH Zurich, FHNW; USA: Caltech, JPL, NHSC.
PACS has been developed by a consortium of institutes led by MPE (Germany) and 
including UVIE (Austria); KU Leuven, CSL, IMEC (Belgium); CEA, LAM (France); 
MPIA (Germany); INAF-IFSI/OAA/OAP/OAT, LENS, SISSA (Italy); IAC (Spain). 
This development has been supported by the funding agencies BMVIT (Austria), 
ESA-PRODEX (Belgium), CEA/CNES (France), DLR (Germany), ASI/INAF (Italy), 
and CICYT/MCYT (Spain).
MT acknowledges support
from MICINN, within the program
CONSOLIDER INGENIO 2010, under grant ``Molecular Astrophysics:
The Herschel and ALMA era - ASTROMOL'' (ref.: CSD2009-00038),
and BN acknowledges support from the ASI project 01/005/11/0.
This research has made use of NASA's Astrophysics Data System 
Bibliographic Services together with the SIMBAD database and 
the VizieR catalogue access tool
operated at CDS, Strasbourg, France.
It also has made use of EURO-VO software, tools, and services. 
The EURO-VO has been funded by the European Commission through 
contracts RI031675 (DCA) and 011892 (VO-TECH) under the 6th Framework 
Programme and contracts 212104 (AIDA) and 261541 (VO-ICE) under 
the 7th Framework Programme.
This work is based in part on observations made with the Spitzer 
Space Telescope, obtained from the NASA/IPAC Infrared Science Archive, 
both of which are operated by the Jet Propulsion Laboratory, 
California Institute of Technology under a contract with the 
National Aeronautics and Space Administration. 
\end{acknowledgements}


\begin{thebibliography}{}



\bibitem[Arce et al.(2007)]{arc07} Arce, H.~G., Shepherd, D., 
Gueth, F., et al.\ 2007, Protostars and Planets V, 245 

\bibitem[Bachiller(1996)]{bac96} Bachiller, R.\ 1996, \araa, 34, 111 

\bibitem[Bachiller et al.(1998)]{bac98} Bachiller, R., Codella, C., Colomer, F., 
Liechti, S., \& Walmsley, C.~M.\ 1998, \aap, 335, 266 

\bibitem[Bachiller et al.(2001)]{bac01} Bachiller, R., P{\'e}rez-Guti{\'e}rrez, M., 
Kumar, M.~S.~N., \& Tafalla, M.\ 2001, \aap, 372, 899 

\bibitem[Beichman et al.(1988)]{bei88} Beichman, C.~A., 
Neugebauer, G., Habing, H.~J., Clegg, P.~E., 
\& Chester, T.~J.\ 1988, Infrared astronomical satellite (IRAS) catalogs and atlases.~Volume 1: Explanatory supplement, 1,

\bibitem[Benedettini et al.(2002)]{ben02} Benedettini, M., Viti, S., 
Giannini, T., Nisini, B., Goldsmith, P.~F., \& Saraceno, P.\ 2002, 
\aap, 395, 657 

\bibitem[Benedettini et al.(2012)]{ben12} Benedettini, M., Busquet, G., 
Lefloch, B., et al.\ 2012, \aap, 539, L3

\bibitem[Bergin et al.(1998)]{ber98} Bergin, E.~A., Neufeld, 
D.~A., \& Melnick, G.~J.\ 1998, \apj, 499, 777

\bibitem[Bergin \& Snell(2002)]{ber02} Bergin, E.~A., \& Snell, R.~L.\ 2002, 
\apjl, 581, L105 

\bibitem[Bergin \& Tafalla(2007)]{ber07} Bergin, E.~A., \& Tafalla, M.\ 2007, 
\araa, 45, 339 

\bibitem[Bieging \& Tafalla(1993)]{bie93}Bieging, J.~H., \& Tafalla, M.\ 1993, 
\aj, 105, 576

\bibitem[Bjerkeli et al.(2009)]{bje09} Bjerkeli, P., 
Liseau, R., Olberg, M., et al.\ 2009, \aap, 507, 1455 

\bibitem[Bjerkeli et al.(2011)]{bje11} Bjerkeli, P., Liseau, R., Nisini, B., 
et al.\ 2011, \aap, 533, A80 

\bibitem[Bjerkeli et al.(2012)]{bje12} Bjerkeli, P., Liseau, R., 
et al.\ 2012, arXiv:1209.0294

\bibitem[Bontemps et al.(1996)]{bon96} Bontemps, S., Andre, P., Terebey, S., 
\& Cabrit, S.\ 1996, \aap, 311, 858 

\bibitem[Bourke et al.(1995)]{bou95} Bourke, T.~L., Hyland, 
A.~R., Robinson, G., James, S.~D., 
\& Wright, C.~M.\ 1995, \mnras, 276, 1067 

\bibitem[Caselli et al.(2012)]{cas12} Caselli, P., Keto, E., 
Bergin, E.~A., et al.\ 2012, \apjl, 759, L37

\bibitem[Castor(1970)]{cas70} Castor, J.~I.\ 1970, \mnras, 149, 111 


\bibitem[Caratti o Garatti et al.(2006)]{car06} Caratti o Garatti, A., 
Giannini, T., Nisini, B., \& Lorenzetti, D.\ 2006, \aap, 449, 1077 

\bibitem[Choi et al.(1999)]{cho99} Choi, M., Panis, J.-F., 
\& Evans, N.~J., II 1999, \apjs, 122, 519 

\bibitem[Codella et al.(2010)]{cod10} Codella, C., Lefloch, B., 
Ceccarelli, C., et al.\ 2010, \aap, 518, L112 

\bibitem[Cunningham et al.(2005)]{cun05} Cunningham, A., 
Frank, A., \& Hartmann, L.\ 2005, \apj, 631, 1010 


\bibitem[Daniel et al.(2010)]{dan10} Daniel, F., Dubernet, M.-L., Pacaud, F., \& Grosjean, A.\ 2010, 
\aap, 517, A13 

\bibitem[Daniel et al.(2011)]{dan11} Daniel, F., Dubernet, M.-L., \& Grosjean, A.\ 2011, \aap, 536, A76 

\bibitem[Draine et al.(1983)]{dra83} Draine, B.~T., Roberge, 
W.~G., \& Dalgarno, A.\ 1983, \apj, 264, 485 

\bibitem[Giannini et al.(2011)]{gia11} Giannini, T., Nisini, 
B., Neufeld, D., Yuan, Y., Antoniucci, S., 
\& Gusdorf, A.\ 2011, \apj, 738, 80 

\bibitem[de Graauw et al.(2010)]{deg10}de Graauw, T., Helmich, F.~P., 
Phillips, T.~G. et al.\ 2010, \aap, 518, L4

\bibitem[Di Francesco et al.(2008)]{dif08} Di Francesco, J., 
Johnstone, D., Kirk, H., MacKenzie, T., 
\& Ledwosinska, E.\ 2008, \apjs, 175, 277

\bibitem[Dionatos et al.(2010a)]{dio10a} Dionatos, O., Nisini, B., Cabrit, S., 
Kristensen, L., \& Pineau Des For{\^e}ts, G.\ 2010, \aap, 521, A7 

\bibitem[Dionatos et al.(2010b)]{dio10b} Dionatos, O., Nisini, B., 
Codella, C., \& Giannini, T.\ 2010, \aap, 523, A29 


\bibitem[Downes \& Cabrit(2003)]{dow03} Downes, T.~P., \& Cabrit, S.\ 2003, \aap, 403, 135 

\bibitem[Dubernet et al.(2006a)]{dub06a} Dubernet, M.-L., Daniel, F., Grosjean, A., et al.\ 2006a, 
\aap, 460, 323 

\bibitem[Dubernet et al.(2006b)]{dub06b} Dubernet, M., Grosjean, A., Daniel, F., et al. 2006b, 
in Ro-vibrational Collisional Excitation Database: BASECOL \url{http://basecol.obspm.fr} (Japan:
Journal of Plasma and Fusion Research Series, series 7)


\bibitem[Dubernet et al.(2009)]{dub09} Dubernet, M.-L., Daniel, F., Grosjean, A., \& Lin, C.~Y.\ 
2009, \aap, 497, 911 

\bibitem[Evans et al.(2009)]{eva09} Evans, N.~J., II, Dunham, 
M.~M., J{\o}rgensen, J.~K., et al.\ 2009, \apjs, 181, 321 

\bibitem[Fazio et al.(2004)]{faz04} Fazio, G.~G., Hora, 
J.~L., Allen, L.~E., et al.\ 2004, \apjs, 154, 10 

\bibitem[Fernandes(2000)]{fer00} Fernandes, A.~J.~L.\ 2000, 
\mnras, 315, 657 

\bibitem[Flower \& Roueff(1999)]{flo99} Flower, D.~R., \& Roueff, E.\ 1999, Journal of Physics B Atomic Molecular Physics, 32, 3399 


\bibitem[Flower \& Pineau Des For{\^e}ts(2010)]{flo10} Flower, D.~R., \& Pineau Des 
For{\^e}ts, G.\ 2010, \mnras, 406, 1745 

\bibitem[Franklin et al.(2008)]{fra08} Franklin, J., Snell, 
R.~L., Kaufman, M.~J., et al.  \apj, 674, 1015 

\bibitem[Frerking et al.(1982)]{fre82} Frerking, M.~A., 
Langer, W.~D., \& Wilson, R.~W.\ 1982, \apj, 262, 590 

\bibitem[Fuente et al.(2005)]{fue05} Fuente, A., Rizzo, J.~R., 
Caselli, P., Bachiller, R., \& Henkel, C.\ 2005, \aap, 433, 535 

\bibitem[Fuller et al.(1995)]{ful95} Fuller, G.~A., Lada, 
E.~A., Masson, C.~R., \& Myers, P.~C.\ 1995, \apj, 453, 754 


\bibitem[Garay et al.(2002)]{gar02} Garay, G., Mardones, D., 
Rodr{\'{\i}}guez, L.~F., Caselli, P., 
\& Bourke, T.~L.\ 2002, \apj, 567, 980 

\bibitem[Gautier et al.(1976)]{gau76} Gautier, T.~N., III, 
Fink, U., Larson, H.~P., \& Treffers, R.~R.\ 1976, \apjl, 207, L129 

\bibitem[Genzel \& Downes(1977)]{gen77} Genzel, R., \& Downes, D.\ 1977, 
\aaps, 30, 145 

\bibitem[Giannini et al.(2001)]{gia01} Giannini, T., Nisini, 
B., \& Lorenzetti, D.\ 2001, \apj, 555, 40 


\bibitem[Goldreich \& Kwan(1974)]{gol74} Goldreich, P., \& Kwan, J.\ 
1974, \apj, 189, 441 


\bibitem[Gredel(1996)]{gre96} Gredel, R.\ 1996, \aap, 305, 582 

\bibitem[Gutermuth et al.(2009)]{gut09} Gutermuth, R.~A., 
Megeath, S.~T., Myers, P.~C., et al.\ 2009, \apjs, 184, 18

\bibitem[Herczeg et al.(2012)]{her12} Herczeg, G.~J., Karska, A., 
Bruderer, S., et al.\ 2012, \aap, 540, A84 

\bibitem[Hirano et al.(2001)]{hir01} Hirano, N., Mikami, H., Umemoto, T., 
Yamamoto, S., \& Taniguchi, Y.\ 2001, \apj, 547, 899 

\bibitem[Hurt \& Barsony(1996)]{hur96} Hurt, R.~L., 
\& Barsony, M.\ 1996, \apjl, 460, L45

\bibitem[Indebetouw et al.(2005)]{ind05} Indebetouw, R., 
Mathis, J.~S., Babler, B.~L., et al.\ 2005, \apj, 619, 931 

\bibitem[Ishihara et al.(2010)]{ish10} Ishihara, D., Onaka, T., Kataza, H., 
et al.\ 2010, \aap, 514, A1

\bibitem[J{\o}rgensen et al.(2004)]{joe04} J{\o}rgensen, J.~K., 
Hogerheijde, M.~R., Blake, G.~A., et al.\ 2004, \aap, 415, 1021 

\bibitem[Kaufman \& Neufeld(1996)]{kau96} Kaufman, M.~J., \& Neufeld, D.~A.\ 
1996, \apj, 456, 611 

\bibitem[Kristensen et al.(2010)]{kri10} Kristensen, L.~E., Visser, R., 
van Dishoeck, E.~F., et al.\ 2010, \aap, 521, L30 

\bibitem[Kristensen et al.(2011)]{kri11} Kristensen, L.~E., 
van Dishoeck, E.~F., Tafalla, M., et al.\ 2011, \aap, 531, L1 

\bibitem[Kristensen et al.(2012)]{kri12} Kristensen, L.~E., van Dishoeck, E.~F., Bergin, E.~A., et al.\ 2012, \aap, 542, A8 

\bibitem[Lefloch et al.(1996)]{lef96} Lefloch, B., Eisloeffel, J., 
\& Lazareff, B.\ 1996, \aap, 313, L17 

\bibitem[Lefloch et al.(2010)]{lef10} Lefloch, B., Cabrit, S., Codella, C., et al.\ 2010, \aap, 518, L113

\bibitem[Lefloch et al.(2012)]{lef12} Lefloch, B., Cabrit, 
S., Busquet, G., et al.\ 2012, \apjl, 757, L25 

\bibitem[Liseau et al.(1996)]{lis96} Liseau, R., Ceccarelli, C., Larsson, B., 
et al.\ 1996, \aap, 315, L181 

\bibitem[Marcaide et al.(1988)]{mar88} Marcaide, J.~M., Torrelles, J.~M., 
Gusten, R., et al.\ 1988, \aap, 197, 235 


\bibitem[Mardones et al.(1997)]{mar97} Mardones, D., Myers, 
P.~C., Tafalla, M., et al.\ 1997, \apj, 489, 719 

\bibitem[Maret et al.(2009)]{mar09} Maret, S., 
Bergin, E.~A., Neufeld, D.~A.  et al.\ 2009, 
\apj, 698, 1244 

\bibitem[McCaughrean et al.(1994)]{mac99} McCaughrean, M.~J., 
Rayner, J.~T., \& Zinnecker, H.\ 1994, \apjl, 436, L189 

\bibitem[Moro-Mart{\'{\i}}n et al.(2001)]{mor01} 
Moro-Mart{\'{\i}}n, A., Noriega-Crespo, A., Molinari, S., et al.\ 2001, 
\apj, 555, 146 

\bibitem[Morris(1976)]{mor76} Morris, M.\ 1976, \apj, 210, 
100 

\bibitem[Mottram et al.(2013)]{mot13} Mottram, J.~C, van Dishoeck, E.~F., 
Kristensen, L.~E., et al.\ 2013, in prep.


\bibitem[Myers \& Ladd(1993)]{mye93} Myers, P.~C., \& Ladd, E.~F.\ 1993, \apjl, 413, L47 

\bibitem[Neufeld \& Dalgarno(1989)]{neu89} Neufeld, D.~A., \& Dalgarno, A.\ 
1989, \apj, 340, 869 


\bibitem[Neufeld \& Yuan(2008)]{neu08} Neufeld, D.~A., \& Yuan, Y.\ 
2008, \apj, 678, 974 

\bibitem[Neufeld et al.(2009)]{neu09} Neufeld, D.~A., 
Nisini, B., Giannini, T.  et al.\ 2009, \apj, 706, 170 

\bibitem[Nisini et al.(1999)]{nis99} Nisini, B., Benedettini, M., 
Giannini, T., et al.\ 1999, \aap, 350, 529

\bibitem[Nisini et al.(2010a)]{nis10a} Nisini, B., Benedettini, M., Codella, C., 
et al.\ 2010a, \aap, 518, L120 

\bibitem[Nisini et al.(2010b)]{nis10b} Nisini, B., Giannini, 
T., Neufeld, D.~A., Yuan, Y., Antoniucci, S., Bergin, E.~A., 
\& Melnick, G.~J.\ 2010b, \apj, 724, 69 

\bibitem[Nisini et al.(2013)]{nis13} Nisini, B., Santangelo, G., 
Antoniucci, S., et al.\ 2013, \aap, 549, A16 

\bibitem[Noriega-Crespo et al.(2004)]{nor04} Noriega-Crespo, 
A., Moro-Mart\'{\i}n, A., Carey, S., et al.\ 2004, \apjs, 154, 402 

\bibitem[Ott(2010)]{ott10} Ott, S.\ 2010, Astronomical Data 
Analysis Software and Systems XIX, 434, 139 

\bibitem[Pickett et al.(1998)]{pic98} Pickett, H.~M., 
Poynter, R.~L., Cohen, E.~A., Delitsky, M.~L., Pearson, J.~C., 
M\"uller, H.~S.~P.\ 1998, \jqsrt, 60, 883 


\bibitem[Pilbratt et al.(2010)]{pil10} Pilbratt, G.~L., Riedinger, J.~R., 
Passvogel, T. et al.\ 2010, \aap, 518, L1

\bibitem[Poglitsch et al.(2010)]{pog10} Poglitsch, A., Waelkens, C., Geis, N. 
et al.\ 2010, \aap, 518, L2

\bibitem[Reach et al.(2006)]{rea06} Reach, W.~T., Rho, J., 
Tappe, A., et al.\ 2006, \aj, 131, 1479 

\bibitem[Rebull et al.(2010)]{reb10} Rebull, L.~M., Padgett, 
D.~L., McCabe, C.-E., et al.\ 2010, \apjs, 186, 259 

\bibitem[Rieke \& Lebofsky(1985)]{rie85} Rieke, G.~H., \& Lebofsky, M.~J.\ 
1985, \apj, 288, 618 

\bibitem[Roelfsema et al.(2012)]{roe12} Roelfsema, P.~R., Helmich, F.~P., 
Teyssier, D., et al.\ 2012, \aap, 537, A17 

\bibitem[Sandell et al.(1994)]{san94} Sandell, G., Knee, L.~B.~G., Aspin, C., 
Robson, I.~E., \& Russell, A.~P.~G.\ 1994, \aap, 285, L1 


\bibitem[Santangelo et al.(2012)]{san12} Santangelo, G., Nisini, B., Giannini, T., 
et al.\ 2012, \aap, 538, A45 

\bibitem[Santiago-Garc{\'{\i}}a et al.(2009)]{san09} Santiago-Garc{\'{\i}}a, 
J., Tafalla, M., Johnstone, D., \& Bachiller, R.\ 2009, \aap, 495, 169 


\bibitem[Sch{\"o}ier et al.(2005)]{sch05} Sch{\"o}ier, F.~L., van der Tak, F.~F.~S., 
van Dishoeck, E.~F., \& Black, J.~H.\ 2005, \aap, 432, 369 

\bibitem[Scoville \& Solomon(1974)]{sco74} Scoville, N.~Z., \& 
Solomon, P.~M.\ 1974, \apjl, 187, L67 

\bibitem[Sep{\'u}lveda et al.(2011)]{sep11} Sep{\'u}lveda, I., Anglada, G., 
Estalella, R., et al.\ 2011, \aap, 527, A41 

\bibitem[Shu(1992)]{shu92} Shu, F.~H.\ 1992, Physics of 
Astrophysics, Vol.~II, (Mill Valley: University Science Books)


\bibitem[Smith et al.(1997)]{smi97} Smith, M.~D., Suttner, G., \& Yorke, H.~W.\ 1997, 
\aap, 323, 223 

\bibitem[Smith et al.(2003)]{smi03} Smith, M.~D., Froebrich, 
D., \& Eisl{\"o}ffel, J.\ 2003, \apj, 592, 245 

\bibitem[Sobolev(1960)]{sob60} Sobolev, V.~V.\ 1960, 
Cambridge: Harvard University Press, 1960,  

\bibitem[Tafalla \& Bachiller(2011)]{taf11} Tafalla, M., \& Bachiller, R.\ 
2011, IAU Symposium, 280, 88 

\bibitem[Tafalla et al.(2000)]{taf00} Tafalla, M., Myers, P.~C., 
Mardones, D., \& Bachiller, R.\ 2000, \aap, 359, 967 

\bibitem[Tafalla et al.(2004)]{taf04} Tafalla, M., Santiago, J., 
Johnstone, D., \& Bachiller, R.\ 2004, \aap, 423, L21 

\bibitem[Tafalla et al.(2006)]{taf06} Tafalla, M., Kumar, M.~S.~N., 
\& Bachiller, R.\ 2006, \aap, 456, 179 

\bibitem[Tafalla et al.(2010)]{taf10} Tafalla, M., Santiago-Garc{\'{\i}}a, J., 
Hacar, A., \& Bachiller, R.\ 2010, \aap, 522, A91 

\bibitem[Valiron et al.(2008)]{val08} Valiron, P., Wernli, 
M., Faure, A., et al.\ 2008, \jcp, 129, 134306 

\bibitem[van Dishoeck \& Blake(1998)]{van98} van Dishoeck, E.~F., \& Blake, 
G.~A.\ 1998, \araa, 36, 317 

\bibitem[van Dishoeck et al.(2011)]{van11} van Dishoeck, E.~F., 
Kristensen, L.~E., Benz, A.~O., et al.\ 2011, \pasp, 123, 138 

\bibitem[Vasta et al.(2012)]{vas12} Vasta, M., Codella, C., Lorenzani, A., 
et al.\ 2012, \aap, 537, A98 

\bibitem[Velusamy et al.(2007)]{vel07} Velusamy, T., Langer, 
W.~D., \& Marsh, K.~A.\ 2007, \apjl, 668, L159 

\bibitem[Velusamy et al.(2011)]{vel11} Velusamy, T., Langer, 
W.~D., Kumar, M.~S.~N., \& Grave, J.~M.~C.\ 2011, \apj, 741, 60 

\bibitem[Yamamura et al.(2010)]{yam10} Yamamura, I., Makiuti, 
S., Ikeda, N., et al.\ 2010, VizieR Online Data Catalog, 2298, 0

\bibitem[White(1977)]{whi77} White, R.~E.\ 1977, \apj, 211, 744 


\bibitem[Wiseman et al.(2001)]{wis01} Wiseman, J., Wootten, 
A., Zinnecker, H., \& McCaughrean, M.\ 2001, \apjl, 550, L87 


\bibitem[Wolniewicz et al.(1998)]{wol98} Wolniewicz, L., 
Simbotin, I., \& Dalgarno, A.\ 1998, \apjs, 115, 293 

\end{thebibliography}
\end{document}